\documentclass[twocolumn,showpacs,preprintnumbers,amsmath,amssymb,superscriptaddress]{revtex4}
\usepackage{epsfig,bm}
\usepackage{ulem}
\usepackage{amsmath}
\begin{document}
\title{Spin-dipole strength functions of $^4$He with realistic nuclear forces}
\author{W. Horiuchi}
\affiliation{Department of Physics, Hokkaido University, Sapporo 060-0810, Japan}
\author{Y. Suzuki}
\affiliation{Department of Physics, Niigata University, Niigata 950-2181, Japan}
\affiliation{RIKEN Nishina Center, Wako 351-0198, Japan}
\pacs{25.10.+s, 21.60.De, 24.30.-v, 27.10.+h}

\begin{abstract}
Both isoscalar and isovector spin-dipole excitations of 
$^4$He are studied using realistic nuclear forces in 
the complex scaling method. The ground state of $^4$He and 
discretized continuum states with $J^{\pi}=0^-,\, 1^-,\, 2^-$ 
for $A=4$ nuclei are described 
in explicitly correlated Gaussians reinforced with global vectors for 
angular motion. Two- and three-body decay channels 
are specifically treated to take into account final state 
interactions. The observed resonance energies and widths of the negative-parity levels are all in fair agreement with those calculated 
from both the spin-dipole and electric-dipole strength functions 
as well as  
the energy eigenvalues of the complex scaled Hamiltonian. 
Spin-dipole sum rules, both non energy-weighted and energy-weighted, 
are discussed in relation to tensor correlations in the 
ground state of $^4$He. 
\end{abstract}
\maketitle

\section{Introduction}

Spin-dipole (SD) excitations of nuclei have attracted much attention
because of their connection with, for example, tensor correlation, neutron
skin-thickness and neutrino-nucleus scattering.
Especially the neutrino reaction involving light nuclei 
is important to the nucleosynthesis at various stages.
In the final stage of a core collapse supernova,
the nuclei are exposed to the intense flux of neutrinos, and  
the neutrino-nucleus reaction rate is determined 
by the nuclear responses
to such operators of the weak interaction as  
Gamow-Teller (GT), dipole, SD, and so on~\cite{gazit07,tsuzuki06}. 
The SD operator brings about the first-forbidden transition 
of the weak interaction.
In the case of $N=Z$ nuclei, the allowed transition probability
due to the weak interaction is small,
and thus the first-forbidden transition can be a leading
order, making a primary contribution to the cross section.

Though the neutrino-nucleus reaction 
cross section can not be measured to good accuracy 
in a laboratory because of  
its too small reaction rate, 
information on the spin excitation of nuclei can be 
obtained using a charge-exchange reaction.
For example, in the $(p,n)$ or $(d,^2$He$)$ reaction 
at intermediate energy, 
the cross section at 0 degree is a useful probe 
to extract the GT strength~\cite{fujita11} as well as the 
SD strength~\cite{okamura02,huu07}. The SD excitation 
can be obtained by measuring the cross section at larger
angles~\cite{gaarde84,rapaport94}. 
For a doubly closed shell nucleus the SD contribution is fairly large
even at 0 degree because of the hindrance of the GT strength.
Much effort has been devoted to measure the SD transitions in light nuclei.
Recently the charge-exchange reaction of ($^7$Li,$^7$Be$\gamma$) is
undertaken to measure the electric dipole $(E1)$ 
and the SD resonances of
$^4$He and $^{6,7}$Li~\cite{nakayama07,nakayama08}.
A more recent measurement of polarization transfer observables with 
$^{16}$O$(\vec{p},\vec{n})^{16}$F reaction~\cite{wakasa11}
indicates that valuable information on the SD excitations is
attainable.

The SD transition is also interesting in 
comparison with the $E1$ transition. 
The SD operator can change the spin wave function of the ground state, 
whereas the $E1$ operator can not. 
Since the SD operator has three possible multipoles, 
the study of its transition strength 
is expected to be more advantageous to see the spin structure of nuclei 
than that by the $E1$ operator~\cite{dumitrescu84}. 
That is, this multipole dependence of the SD operator may 
be used to probe the role of non-central forces,
especially the tensor force. 
The effect of the tensor force has in fact been studied theoretically
by looking at the SD excitations in the shell model~\cite{tsuzuki98} and in the 
random phase approximation based on 
Skyrme-Hartree-Fock~\cite{bai10,bai11} and relativistic
Hartree-Fock methods~\cite{liang12}.
All these calculations employed the variety of effective interactions and
found that the residual tensor terms added to the interaction play
some multipole-dependent effects on the SD strength functions.

The purpose of this paper is to study the SD excitations of $^4$He 
using realistic nucleon-nucleon interactions. Only the ground state of 
$^4$He is bound among $A=4$ nucleon systems and its basic property 
is now understood fairly well thanks to several accurate methods for 
solving bound state problems of few-body systems~\cite{kamada01}. 
All the excited states of $^4$He are in continuum and its 
negative-parity states below the $2n+2p$ threshold 
have $J=0, 1, 2$ with $T=0$ and 1. These resonances 
as well as the continuum states may be reached by the SD operators. 
It is therefore quite challenging to accurately predict the SD strength 
function as a function of excitation energy because we have to deal with 
the continuum states where not only two but also three particles may play 
an important contribution. On top of that we have to take into account 
both short-range and tensor correlations due to the realistic nuclear
forces~\cite{forest96,feldmeier11}. Very recently the present authors and Arai 
have done an {\it ab initio} calculation 
for the photoabsorption of $^{4}$He~\cite{horiuchi12a} using square integrable 
($\mathcal{L}^2$) basis functions in the framework of 
the complex scaling method (CSM) and have 
reproduced most of experimental photoabsorption cross section data 
up to the pion threshold. A theoretical approach employed in the
present paper is similar to that of the $E1$ case. 

It is well known that the ground state of $^4$He contains 
the $D$-state (or the total spin $S=2$ state) probability by about 14\%, 
which is of course due to the tensor force. 
As shown in the calculation of bound-state 
approximation~\cite{horiuchi08,horiuchi12b},
the tensor force plays a vital role in correctly reproducing the
spectrum of the excited states of $^4$He. 
If one uses such effective interactions that contain no tensor
components, there is no way to account for the level splittings of the 
negative-parity states of $^4$He. Therefore the use of realistic nuclear 
forces is absolutely necessary for studying the SD strength in
$^4$He. In the same context we also study the charge-exchange 
SD transitions from $^4$He, leading to the negative-parity states of 
$^4$H or $^4$Li. We will pay due attention to the effect of the tensor 
correlation on the SD excitations. It should be noted that the
SD excitation is here described based on the accurate ground-state wave
function of $^4$He~\cite{horiuchi08,horiuchi12b}. We also note that 
this study will serve fundamental data on  
the neutrino-$^4$He reaction cross section in stars by integrating 
the SD strength functions weighted by the neutrino energy distribution 
produced by the  core collapse star.

In Sec.~\ref{method.sec} we present our method of
evaluating the SD strength functions, the CSM (Sec.~\ref{CSM.sec})
and the $\mathcal{L}^2$ basis functions (Sec.~\ref{model.sec}) that are keys in the present 
calculation. We show calculated results
on the SD strength functions in Sec.~\ref{results.sec}. The SD
strengths calculated from continuum-discretized states are presented
in Sec.~\ref{discretized.strength}. The SD strength functions of both 
isovector and isoscalar types 
are displayed in Sec.~\ref{SDstrength.f}. A comparison of the peaks of
the SD strength functions with the resonance properties of $^4$He is made in 
Sec.~\ref{resonance.para}. The SD sum rules, both non energy-weighted 
and energy-weighted, are discussed in Sec.~\ref{sec.SDSR}. 
Conclusions are drawn in Sec.~\ref{conclusions.sec}. 
A multipole decomposition of the SD non energy-weighted sum
rule (NEWSR) is 
discussed in Appendix A, and a method of calculating its relevant matrix
element with our basis functions 
is briefly explained in Appendix B. In Appendix C we derive a 
formula that makes it possible to calculate the contribution of the
kinetic energy operator to the SD energy-weighted sum rule (EWSR). 

\section{Calculation method of spin-dipole strength function}
\label{method.sec}

\subsection{Complex scaling method}
\label{CSM.sec}

The SD operator with the multipolarity $\lambda$ and its projection $\mu$ is defined by
\begin{align}
\mathcal{O}_{\lambda\mu}^{p}=
\sum_{i=1}^{N}\left[\bm{\rho}_i\times\bm{\sigma}_i\right]_{\lambda\mu}
 T^p_i
\label{sd.op}
\end{align}
with 
\begin{align}
\bm \rho_i=\bm{r}_i-\bm{x}_{N},
\end{align}
where $\bm{r}_i$ is $i$th nucleon coordinate, $\bm{x}_{N}$ is the center-of-mass 
coordinate of the ${N}$-nucleon system, and $\bm \sigma_i$ is 
$i$th nucleon spin. 
The center-of-mass motion is completely removed in the 
present paper and only the intrinsic excitation is considered. 
The square bracket 
$[\bm{\rho}_i \times \bm{\sigma}_i]_{\lambda \mu}$ denotes 
the angular momentum coupling of the two vectors or more
generally the tensor product of spherical tensors to  
that operator specified by $\lambda\, \mu$.  
The value of $\lambda$ can take 0, 1, and 2.
The superscript $p$ of $\mathcal{O}_{\lambda\mu}^{p}$ or
$T^p_i$ distinguishes different types of isospin operators, 
isoscalar (IS), isovector (IV0),
and charge-exchange (IV$+$ and IV$-$), that is,
\begin{align}
&T^{\rm IS}_i=1,\quad T^{\rm IV0}_i=\tau_z(i),\quad 
T^{\rm IV\pm}_i=t_{\pm}(i).
\end{align}
In the inelastic neutrino-nucleus reaction,
the neutral current induces the IV0 type operator 
as well as the IS one. 
The isospin operator $t_{+}=t_x + it_y$  $(t_{-}=t_x-it_y)$ 
converts a proton (neutron) to a neutron (proton), which corresponds
to the charge-exchange process $X(n,p)Y$ $(X(p,n)Y)$.

The strength function of an initial state 
$\Psi_0$ for the SD operator is defined as
\begin{align}
&S(p,\lambda,E)=\mathcal{S}_{f\mu}
|\left<\Psi_f\right|\mathcal{O}^p_{\lambda\mu}\left|\Psi_0\right>|^2
\delta(E_f-E_0-E)\notag\\
&=-\frac{1}{\pi}\text{Im}\sum_{\mu}\left<\Psi_0\right|
\mathcal{O}^{p\dagger}_{\lambda\mu}\frac{1}{E+E_0-H+i\epsilon}
\mathcal{O}^p_{\lambda\mu}
\left|\Psi_0\right>,
\label{sf.eq}
\end{align}
where $\mathcal{S}_{f\mu}$ represents 
a summation over $\mu$ and all the final states $\Psi_f$.  
Both the initial and final states are the  
eigenfunctions of a Hamiltonian $H$ with the energies  
$E_0$ and $E_f$. They are normalized as usual:   
$\left<\Psi_{\nu^\prime}|\Psi_\nu\right>=\delta_{\nu^\prime\nu}$ and 
$\delta(E_{\nu^\prime}-E_\nu)$ for bound and unbound states, 
respectively. In the second expression of Eq.~(\ref{sf.eq}) 
the summation over the final states with the energy conservation of 
$\delta(E_f-E_0-E)$  is converted to the imaginary part of a 
resolvent $R$
\begin{align}
R=\frac{1}{E+E_0-H+i\epsilon}.
\end{align}

In the present paper we use the CSM to obtain the strength function. The CSM  
is widely used not only in atomic and molecular physics~\cite{ho83,moiseyev98} 
but in nuclear physics~\cite{CSM} as well. 
Very recently it has successfully been applied to calculate 
the photoabsorption cross section of $^4$He 
with a realistic Hamiltonian~\cite{horiuchi12a}. 
The CSM allows us to obtain the strength function using only $\mathcal{L}^2$
basis functions exclusively, making it possible to avoid an explicit
construction of the continuum state.   
The key of the CSM is to rotate both the coordinate and the momentum by 
a scaling angle $\theta$
\begin{align}
\bm{r}_j \to \bm{r}_je^{i\theta},\quad 
\bm{p}_j \to \bm{p}_je^{-i\theta},
\label{trans.eq}
\end{align}
which makes the continuum state damp at large distances within 
a certain range of $\theta$.
The strength function $S(p,\lambda,E)$ reduces to
\begin{align}
&S(p,\lambda,E)\notag \\
&=-\frac{1}{\pi}\text{Im}\sum_\mu\left<\Psi_0\right|
\mathcal{O}_{\lambda\mu}^{p\dagger} U^{-1}(\theta)R(\theta)
U(\theta)\mathcal{O}^p_{\lambda\mu}\left|\Psi_0\right>,
\end{align}
where $U(\theta)$ is the scaling operator that makes the transformation (\ref{trans.eq}) 
and $R(\theta)$ is the complex scaled  
resolvent
\begin{align}
R(\theta)=U(\theta)RU^{-1}(\theta)=\frac{1}{E+E_0-H(\theta)+i\epsilon}
\end{align}
with the rotated Hamiltonian
\begin{align}
H(\theta)=U(\theta)HU^{-1}(\theta).
\end{align}
Provided the eigenfunctions of $H(\theta)$ 
are made to damp at large distances, they can be 
expanded with a set of 
$\mathcal{L}^2$ basis functions $\Phi_i (\bm{x})$
\begin{align}
\Psi_\nu(\theta)=\sum_iC_i^\nu (\theta)\Phi_i(\bm{x}).
\end{align}
The coefficients $C_i^\nu (\theta)$ together with the complex 
eigenvalue $E_\nu(\theta)$ are determined by 
diagonalizing $H(\theta)$:
\begin{align}
H(\theta)\Psi_\nu(\theta)=E_\nu(\theta)\Psi_\nu(\theta).
\label{ceig.eq}
\end{align}
The strength function $S(p,\lambda,E)$ is then calculated from the following expression: 
\begin{align}
S(p,\lambda,E)=-\frac{1}{\pi}\sum_{\nu,\mu} \text{Im}
\frac{\tilde{\mathcal{D}}^{p,\nu}_{\lambda\mu}(\theta)\mathcal{D}^{p,\nu}_{\lambda\mu}(\theta)}
{E+E_0-E_\nu(\theta)+i\epsilon},
\label{spectr.exp}
\end{align}
where 
\begin{align}
&\mathcal{D}^{p,\nu}_{\lambda\mu}(\theta)=\left<(\Psi_\nu(\theta))^* 
\right|\mathcal{O}^p_{\lambda\mu}(\theta)\left|U(\theta)\Psi_0 \right>,\notag\\
&\mathcal{\tilde{D}}_{\lambda\mu}^{p,\nu}(\theta)=\left<(U(\theta)\Psi_0)^* 
\right|\tilde{\mathcal{O}}^p_{\lambda\mu}(\theta)\left|\Psi_\nu(\theta)\right>
\end{align}
with
\begin{align}
\mathcal{O}_{\lambda\mu}^p(\theta)=\mathcal{O}_{\lambda\mu}^p
e^{i\theta},\quad
\tilde{\mathcal{O}}_{\lambda\mu}^p (\theta)=\mathcal{O}_{\lambda\mu}^{p\dagger}e^{i\theta}.
\end{align}
Note that $U(\theta)\Psi_0$ is here taken to be 
the solution of Eq. (\ref{ceig.eq})
corresponding to the initial state. 

If a sharp resonance exists, 
the angle $\theta$ has to be rotated to cover its resonance pole 
on the complex energy plane~\cite{moiseyev98,CSM}. 
Practically the scaling angle $\theta$ is chosen by examining 
the stability of the strength function with respect to $\theta$. 
See Refs.~\cite{horiuchi12a,horiuchi12b} for some examples on the 
$\theta$-dependence. 

\subsection{Correlated Gaussians and global vectors}

\label{model.sec}

\subsubsection{Hamiltonian}

The Hamiltonian $H$ we use contains 
two- and three-nucleon interactions
\begin{align}
H=\sum_{i=1}^{N} T_i-T_\text{cm}+\sum_{i<j}v_{ij}+\sum_{i<j<k}v_{ijk}.
\end{align}
In the kinetic energy $T_i$ the proton-neutron mass difference is ignored. 
Two different two-nucleon interactions, 
AV8$^\prime$~\cite{AV8p} and G3RS~\cite{G3RS} potentials, 
are employed to examine the extent to which the strength function 
is sensitive to the $D$-state
probability of $^4$He. The $\bm{L}^2$ and $(\bm{L}\cdot\bm{S})^2$ terms
 in the G3RS potential are ignored. 
The AV8$^\prime$ potential is more  
repulsive at short distances
and has a stronger tensor component than the G3RS potential. 
As the three-body interaction (3NF) we adopt the spin-isospin independent 
phenomenological potential~\cite{hiyama04} that is 
adjusted to reproduce both the  inelastic electron scattering form 
factor to the first excited state of $^{4}$He as well as the binding energies of $^{3,4}$He and $^3$H. 
The Coulomb potential is included, 
but the isospin is treated as a conserved quantum number. The
nucleon mass $m_N$ and the charge constant $e$ used in what follows  
are $\hbar^2/m_N=41.47106$ MeV\,fm$^2$ and $e^2=1.440$ MeV\,fm.

\subsubsection{Basis functions for bound states}

We solve the four-body Schr\"{o}dinger equation using a variational method.
A choice of the variational trial functions is essential to determine
the accuracy of the calculation.
A bound-state solution with spin-parity $J^\pi$ of $N$-nucleon system 
may be expressed in terms of a linear combination of the $LS$ coupled 
basis functions
\begin{align}
\Phi_{(LS)JM_JTM_T}^{\pi}=\mathcal{A}\left[\phi_L^{\pi}
\times\chi_S\right]_{JM_J}
\eta_{TM_T},
\label{lscoupled}
\end{align}
where $\mathcal{A}$ is the antisymmetrizer, and  
the spin function $\chi_S$ is given  
in a successive coupling as 
\begin{align}
&\chi_{S_{12},S_{123},\dots, S M_S}\notag\\
&=[\dots[[\chi_\frac{1}{2}(1)\times\chi_\frac{1}{2}(2)]_{S_{12}}
\times\chi_\frac{1}{2}(3)]_{S_{123}}\dots]_{SM_S}.
\label{spin.fn}
\end{align}
Note that the above spin function forms 
a complete set provided all
possible intermediate spins ($S_{12}, S_{123}$, $\dots$) 
are included for a given $S$. 
The isospin function $\eta_{TM_T}$  
is given in exactly the same way as the spin function.

The spatial part $\phi_L^{\pi}$ 
should be flexible enough to cope with the strong tensor force
and short-range repulsion. 
The tensor force mixes the $S$ and $D$ components 
in the wave function and the short-range repulsion 
makes the amplitude of the two-nucleon 
relative motion function vanishingly small at short distances.  
Many examples show that the correlated Gaussian (CG) basis~\cite{boys60,singer60} is flexible enough to meet these requirements
~\cite{varga97,kamada01,horiuchi08}.  
See a recent review~\cite{mitroy13} for various powerful 
applications of the CG. 
Let an $(N-1)$-dimensional column vector or an $(N-1)\times 1$ matrix 
$\bm{x}$ denote a set of relative coordinates whose $i$th 
element is a  3-dimensional vector $\bm{x}_i$. 
A set of the Jacobi coordinates is most often employed for $\bm x$ but
other sets of relative coordinates may be used as well.  
The spatial part $\phi_{L}^{\pi}$, given in
the CG with two global vectors (GV), takes a 
form~\cite{varga95,svm,suzuki08,aoyama12}
\begin{align}
&F_{(L_1L_2)LM_L}(u_1, u_2, A, \bm{x})\notag\\
&=\exp(-\textstyle{\frac{1}{2}}\tilde{\bm{x}}A\bm{x})
\left[\mathcal{Y}_{L_1}(\tilde{u}_1\bm{x})\times\mathcal{Y}_{L_2}(\tilde{u}_2\bm{x})\right]_{LM_L}
\label{GVR.eq}
\end{align}
with 
\begin{align}
\mathcal{Y}_{\ell m}(\bm v)=v^{\ell}Y_{\ell m}(\hat{\bm v}),
\end{align}
where $A$ is an $(N-1)\times (N-1)$ positive-definite, symmetric matrix
and $\tilde{\bm{x}}A\bm{x}$ is a short-hand notation for
$\sum_{i,j=1}^{N-1}A_{ij}\bm{x}_i\cdot\bm{x}_j$.
The tilde stands for the transpose of a matrix.
Parameters $u_1$ and $u_2$ are $(N-1)$-dimensional column vectors that 
define the GVs, $\tilde{u}_1\bm x(=\sum_{i=1}^{N-1}{u_1}_i\bm x_i)$ and $\tilde{u}_2\bm x$, and these characterize the angular motion of the system.

The CG-GV basis~(\ref{GVR.eq}) apparently describes correlated motion 
among the particles
through the off-diagonal elements of $A$ and the rotational motion 
of the system is conveniently 
described by different sets of $(L_1, L_2)$ carried by the two GVs. 
Most noticeable among several 
advantages of the CG-GV basis functions are that 
the functional form of Eq.~(\ref{GVR.eq}) remains unchanged 
under an arbitrary linear 
transformation of the coordinate $\bm x$, that the matrix elements for most operators can be evaluated analytically, and that 
the formulation can readily be extended to systems with larger $N$. 
Useful formulas for evaluating matrix elements with the CG-GV basis 
are collected in Appendices of Refs.~\cite{suzuki08,aoyama12}.

All possible $L,\, S$ sets are adopted to specify the basis functions 
for a given $J$. The value of $S$
can be 0, 1, and 2 for the four-nucleon system, and all possible $L$ values
that make $J$ with $S$ are included.  
For a given $L^{\pi}$ we choose the simplest combination 
of ($L_1$,\, $L_2$): ($L_1=L,\, L_2=0$) for a natural parity state with 
$\pi=(-1)^L$ and $(L_1=L,\, L_2=1$) for an unnatural parity state with 
$\pi=(-1)^{L+1}$, respectively. An exception is that 
no basis function with $L^\pi=0^-$ is included in our
calculation because that special case 
needs at least three GVs~\cite{suzuki00,aoyama12}. It should be noted,
however, that the $L^{\pi}=0^-$ configuration may be excited by the 
$E1$ and SD transitions  
only through the $(L=S=1)$ component of the ground state of $^4$He.  
Since the probability of finding that component is quite small 
(less than 0.4\%)~\cite{kamada01,suzuki08},
practically we do not miss any SD strength by the neglect of 
the $L^{\pi}=0^-$ configuration. This is really the case in  
the $E1$ strength function~\cite{horiuchi12a} and in the 
SD case as well as shown in Sec.~\ref{sec.SDSR}.

The parameters, $A$, $u_1$, and $u_2$, are determined by the 
stochastic variational method (SVM)~\cite{varga94,varga95,svm}.  
The calculated properties of $^3$H, $^3$He, and $^4$He 
agree with experimental three- and four-nucleon data
very well~\cite{horiuchi12b}. 

\subsubsection{Square-integrable basis functions for spin-dipole excitations}

\label{wavefn.sec}

We construct the basis functions for the final states
with $J^-T$ 
that are excited by the SD operator $\mathcal{O}_{\lambda\mu}^{p}$ 
of $\lambda=J$. 
The accuracy of the CSM calculation depends on how 
fully the basis functions are prepared.
In Ref.~\cite{horiuchi12a}, the present authors
and Arai described a way to construct  
the four-body continuum-discretized states with $J^{\pi}T=1^-1$.  
The guidelines of the construction were to 
take into account both sum rule and final state interactions 
between the particles in the continuum.  
The total photoabsorption cross section is calculated
via the $E1$ strength function and it succeeds to 
reproduce the measured cross section up to the pion threshold. 
Here we take the same route as that of 
Ref.~\cite{horiuchi12a} with 
a possible modification due to the spin flip of
the SD operator.

We define a single-particle (sp) basis,
which describes a single-particle like excitation from the correlated 
ground state by the SD operator. This class of basis functions is 
expected to play a vital role in accounting for all the SD strength.  
The basis is constructed as 
\begin{align}
\Psi_f^{\text{sp}}=\mathcal{A}\left[\left[\phi_{L_i}^{+}(i)\times\chi_{S^\prime}\right]_{J^\prime}\times
\mathcal{Y}_1({\bm{\rho}}_1)\right]_{\lambda \mu}\eta_{TM_T},
\end{align}
where $\phi_{L_i}^{+}(i)$ is the space
part of the $i$th basis function of a 
truncated ground-state wave function, $\Psi_{0000}^{+}$, of $^4$He.
The wave function $\Psi_{0000}^{+}$ consists of 
$[\phi_0^{+}\times\chi_0]_{00}\eta_{00}$ and  
$[\phi_2^{+}\times\chi_2]_{00}\eta_{00}$, with 
any configurations
of $[\phi_1^{+}\times\chi_1]_{00}\eta_{00}$ being omitted, 
which leads to  
1.53 MeV loss for the ground-state energy of $^4$He. 
See Ref.~\cite{horiuchi12a} for the detail. 
As for the spin part, differently from the $E1$ case~\cite{horiuchi12a} 
we take into account the complete set for a given $S^\prime$,  
which, depending on the total spin $S_i$ of the $i$th 
basis function of $\Psi_{0000}^{+}$, is chosen as $S^\prime=1$ for $S_i=0$ and 
$S^\prime=1$ and 2 for $S_i=2$, respectively. 

The $3N+N$ two-body and $d+p+n$ three-body disintegration channels
are defined in the same manner as in Ref.~\cite{horiuchi12a}. 

The calculations are performed not only in each basis set of  
sp, $3N+N$ and $d+p+n$ but also in 
the `Full' basis that includes all of
them. The number of basis functions in the Full model with the 
AV8$^\prime$+3NF potential is 
2980, 6400, 6540, 4380, 8800, 9540 for $J^\pi T=0^-0, 1^-0, 2^-0, 
0^-1, 1^-1, 2^-1$, respectively.

\section{Results and discussions}
\label{results.sec}

\subsection{Discretized spin-dipole strength}
\label{discretized.strength}

\begin{figure*}[ht]
\begin{center}
\epsfig{file=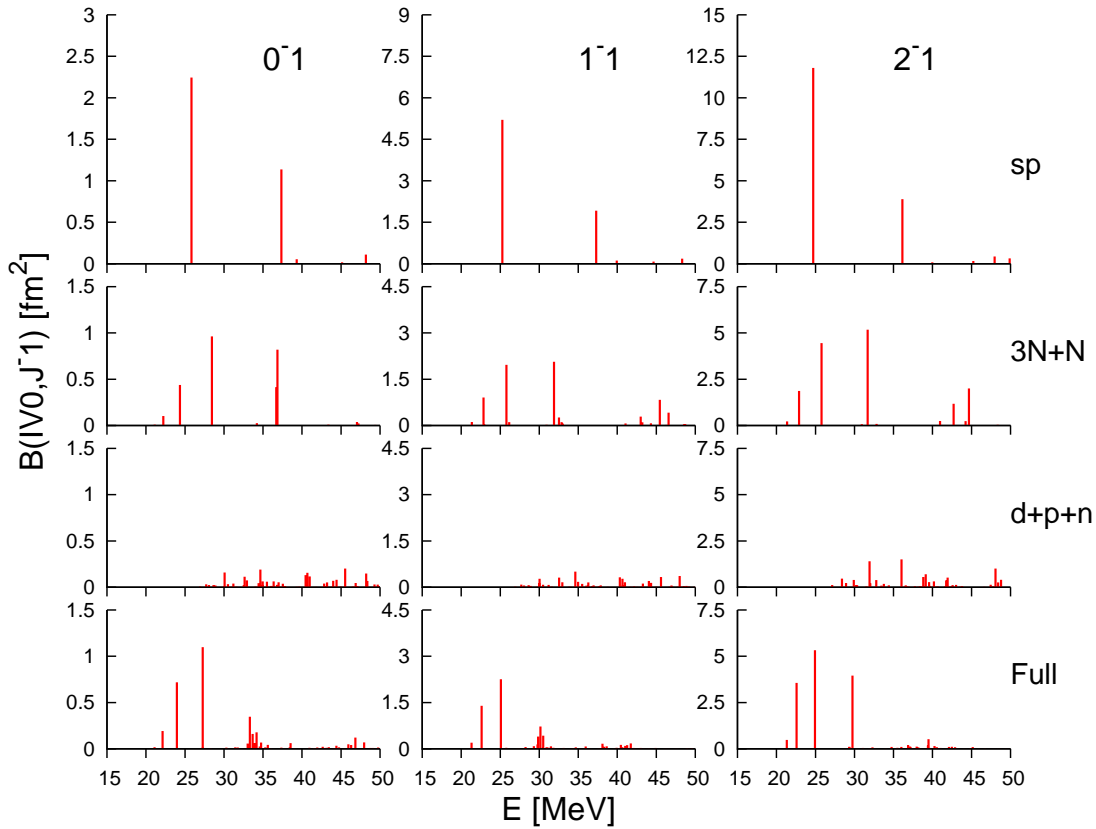,scale=1.1}
\caption{(Color online) The discretized isovector SD reduced transition
 probabilities of IV0 type for $^4$He as a function of excitation
 energy. The $J^{\pi}T$ values of the excited states are $0^-1, 1^-1,$
 and $2^-1$, respectively. The transition probabilities are displayed, 
 arranged vertically for each $J^{\pi}T$ case, 
 depending on the configurations included in the calculation. 
See text for details.
The AV8$^\prime$+3NF interaction is used.}
\label{SDIVdisc.fig}
\end{center}
\end{figure*}

First we show the continuum-discretized SD strength. 
For this purpose the Hamiltonian is diagonalized in the basis states that are 
defined in Sec.~\ref{wavefn.sec}. This calculation 
corresponds to the CSM solution with $\theta=0^\circ$.
Figure~\ref{SDIVdisc.fig} displays the reduced transition probability
for the IV SD operator ($p={\rm IV0}$)
\begin{align}
B(p,J^-T,\nu)=\sum_{M\mu}\left|\right<\Psi^{J^-MT}_\nu(\theta=0^\circ)|
\mathcal{O}_{\lambda\mu}^p
|\Psi_{0}\left>\right|^2,
\end{align}
as a function of the discretized energy $E_\nu (\theta=0^\circ)$,
where $\nu$ is the label to distinguish the discretized energy. 
Here $\lambda$ is equal to $J$ and $T=1$. 

The results of calculation 
are similar to the $E1$ case of Ref.~\cite{horiuchi12a}.
In the calculation  with the sp configuration only,
the strength is concentrated at one state that appears 
at about 25 MeV for all the cases with 
$J^\pi T=0^-1,\, 1^-1,\, 2^-1$. The state may  
correspond to the observed level at 
23.33 MeV for $J^\pi T=2^-1$ and 25.28 MeV for $J^\pi T=0^-1$, 
respectively~\cite{tilley92}.
For the $J^\pi T=1^-1$ case, two levels with very broad widths are known 
at 23.64 and 25.95 MeV. Since the energy of the prominent SD transition 
strength is lower than that obtained for the $E1$ transition 
strength~\cite{horiuchi12a}, the 
23.64 MeV level probably has SD character, whereas the 25.95 MeV 
level is excited by the $E1$ operator.  

Similarly to the $E1$ transition strength~\cite{horiuchi12a},
two or three peaks are obtained with the $3N+N$ configuration
and relatively small strength is spread broadly above 30 MeV.
The prominent peaks below 30 MeV shown in   
the $3N+N$ calculation continue to remain  
in the Full basis calculation, which again confirms the importance of
the $3N+N$ configuration to describe the low-energy SD
strengths as in the $E1$ strength.
We also calculate the SD strength with the G3RS+3NF interaction. 
Both distributions look similar, indicating the weak 
dependence of the SD strength on the realistic interactions employed.

The so-called softening and hardening of the SD
excitation is discussed in Refs.~\cite{bai10,bai11}, where 
the residual tensor force is turned on or off 
and the energy and the
strength of the SD excitation are compared each other. 
We think that the conclusion drawn by such 
comparisons is not always true because switching off the important 
piece of the nucleon-nucleon interaction may cause a significant change 
in the continuum structure that can be reached by the SD 
operator. In fact, we can not turn off the tensor force. If the 
tensor force were switched off, the ground state of $^4$He would not be
bound and moreover the spectrum of the negative-parity  
states would be far from the observed one~\cite{horiuchi08,horiuchi12b}. 
As will be discussed later, 
we find no quenching of the SD strength but confirm that 
our SD strength calculated with  
the realistic nuclear forces satisfies the NEWSR perfectly. 

\begin{figure*}[ht]
\begin{center}
\epsfig{file=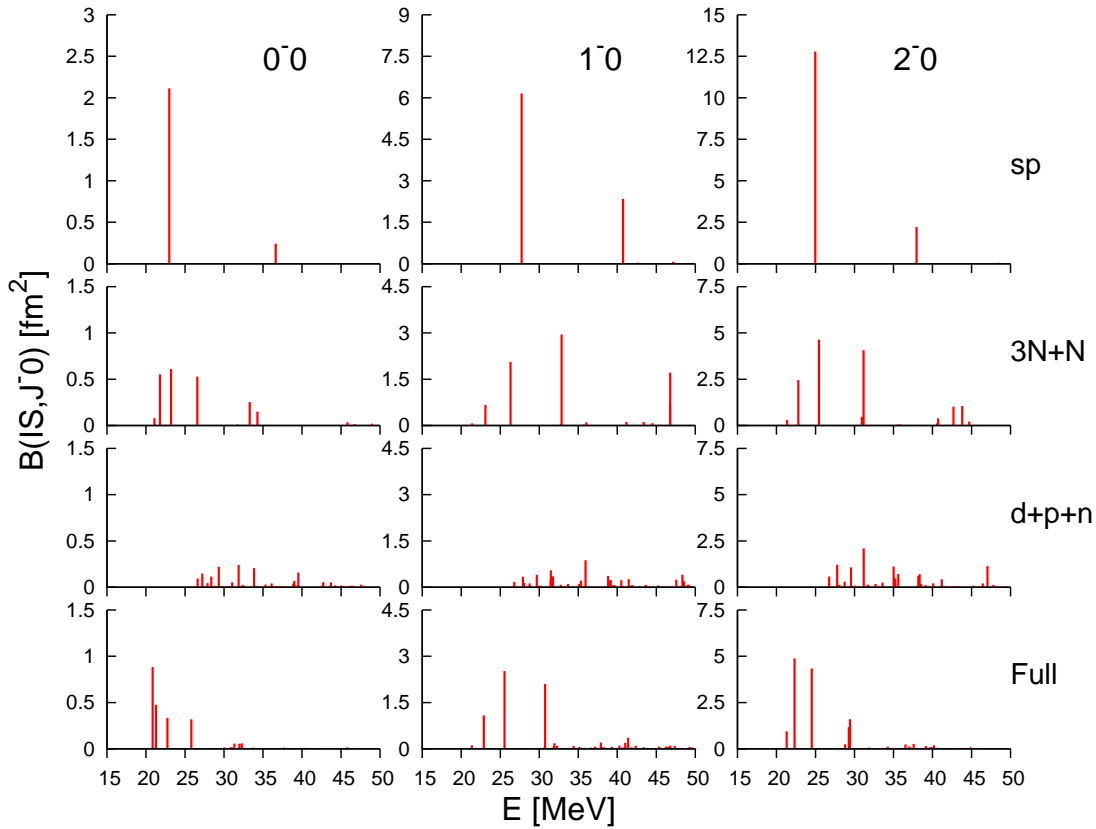,scale=1.1}
\caption{(Color online) The same as
 Fig.~\ref{SDIVdisc.fig} but for the isoscalar SD transitions.
The $J^{\pi}T$ values of the excited states are $0^-0, 1^-0,$
 and $2^-0$, respectively.}
\label{SDISdisc.fig}
\end{center}
\end{figure*}

Figure~\ref{SDISdisc.fig} displays
the reduced transition probability for the IS SD operator 
($p={\rm IS}$). The gross structure of the strength is similar to the 
IV SD case. In the sp configuration calculation, strongly 
concentrated peaks appear at the energies not far from the 
observed levels~\cite{tilley92}: 21.01, 21.84, 24.25 MeV for 
$J^\pi T=0^-0, 2^-0, 1^-0$, respectively. 
The importance of the $3N+N$ configurations 
is indicated by those peaks that appear in the $3N+N$ 
configuration calculation and continue to exist in the Full calculation.
The Full calculation predicts one prominent peak at 20.85 MeV
for $J^\pi T=0^-0$, which may correspond to the 21.01 MeV level with 
the small decay width of $0.84$ MeV~\cite{tilley92}. In the case of $J^\pi
T=2^-0$, the two prominent strengths are obtained at the energies 
close to each other, suggesting a relatively small decay width. The
relationship between the strength functions and resonance properties
will be discussed in Sec.~\ref{resonance.para}.

Three negative-parity states of $^4$He 
with $J=0,\, 1,\, 2$ and $T=0$ are observed slightly above the 
$2n+2p$ threshold of 28.3\,MeV~\cite{tilley92}. 
These states may be excited by
the IS SD operator.  
In fact, a comparison of the IS and IV SD strengths 
obtained in the $d+p+n$ configurations clearly suggests that more
strength is found in the IS case around the excitation energy of 30
MeV. Therefore some discretized states at around 30 MeV shown in
Fig.~\ref{SDISdisc.fig} may be precursors of those observed states. 
Since the three observed states almost entirely 
decay by emitting deuterons, it is likely that 
they have $d+d$ structure with a $P$-wave relative motion. 
The $P$-wave relative motion 
is possible only when the channel spin of two $d$s is coupled to 
1~\cite{aoyama12}. Thus 
we have the possibility of $J^{\pi}=0^-,\, 1^-$, and $2^-$ in accordance
with the observation. Our basis functions partially include 
the $d+d$ type configurations, but an explicit inclusion of them may  
be desirable to discuss this issue in more detail.

\begin{figure*}[ht]
\begin{center}
\epsfig{file=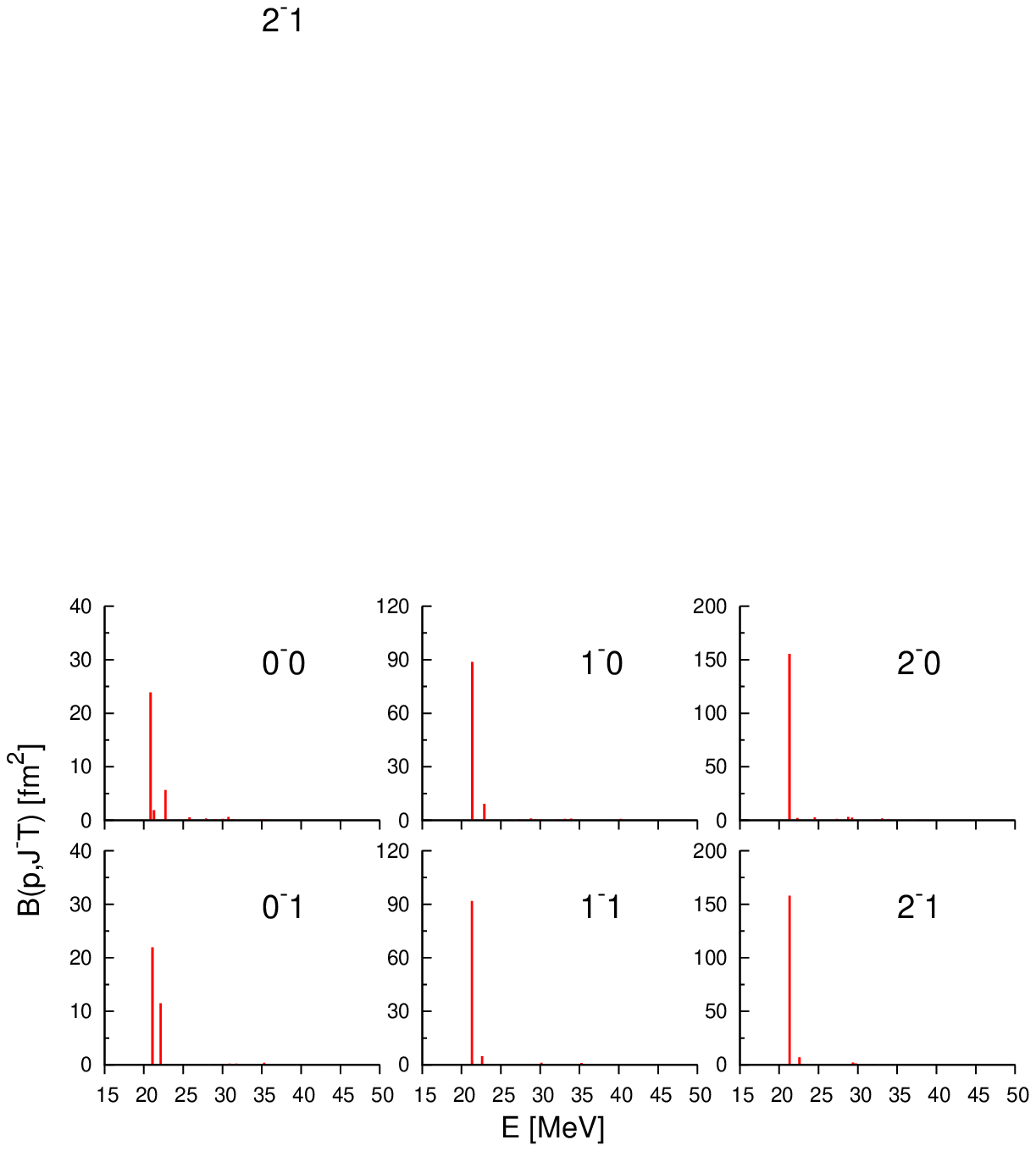,scale=1.1}
\caption{(Color online) 
The discretized SD reduced transition
probabilities of  IS and IV0 types from the first excited $0^+0$ state
of $^4$He as a function of excitation energy. The calculation is
 done in the Full basis using the AV8$^{\prime}$+3NF interaction. 
The calculated excitation energy of the $0^+0$ state is 20.33 MeV 
from the ground-state of $^4$He~\cite{horiuchi12b}.}
\label{SDdisc-2nd.fig}
\end{center}
\end{figure*}

It is interesting to recall the enhanced SD excitations of the first excited
state $J^\pi T=0_2^+0$ that is interpreted as
having a well developed $3N+N$ ($^3$H+$p$ and $^3$He+$n$) 
structure~\cite{hiyama04,horiuchi08}.
It is shown in Ref.~\cite{horiuchi08} that some 
negative-parity states can also be  
understood as parity inverted partners of the first excited state
and the SD transition strengths from that state
are quite enhanced and mostly exhausted by 
only those negative-parity states.
Figure~\ref{SDdisc-2nd.fig} exhibits 
the SD reduced transition probabilities 
from the $0_2^+0$ state as a function of excitation energy.
The transition probabilities of both IS and IV0 are very much 
enhanced, approximately 20-30 times
larger than those from the ground state and each of the strengths  
is concentrated at the respective peak. The excitation 
energies of the peaks are 
20.85, 21.37, 21.30 MeV for $J^\pi T=0^-0, 1^-0, 2^-0$
and 21.10, 21.32, 21.33 MeV for $J^\pi T=0^-1, 1^-1, 2^-1$, respectively.
The energy required for the $0_2^+0$ state to reach the peak position 
is only $0.5-1.0$ MeV. 
The neutrino reaction rate would be greatly enhanced 
if there were such a situation in which 
a plenty of the first excited states of $^4$He existed in 
the core collapse star. 
The situation may, however, be unlikely as the 
life time of that state is short
and its excitation energy (20.21 MeV) is considerably high  
compared to the typical temperature 
of the collapsing star~\cite{woosely90}.

\subsection{Spin-dipole strength functions}

\label{SDstrength.f}

\begin{figure}[ht]
\begin{center}
\epsfig{file=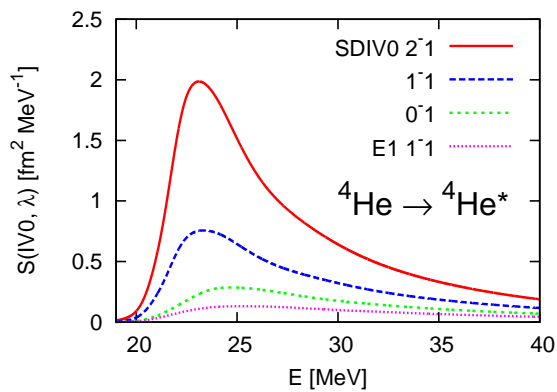,scale=1.2}
\caption{(Color online) Isovector SD strength functions
 of IV0 type and $E1$ strength function for $^4$He
 as a function of excitation energy.}
\label{SD-IV.fig}
\end{center}
\end{figure}

\begin{figure}[ht]
\begin{center}
\epsfig{file=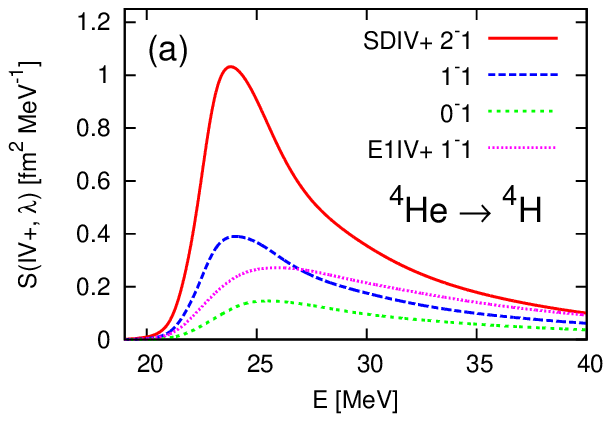,scale=1.2}
\epsfig{file=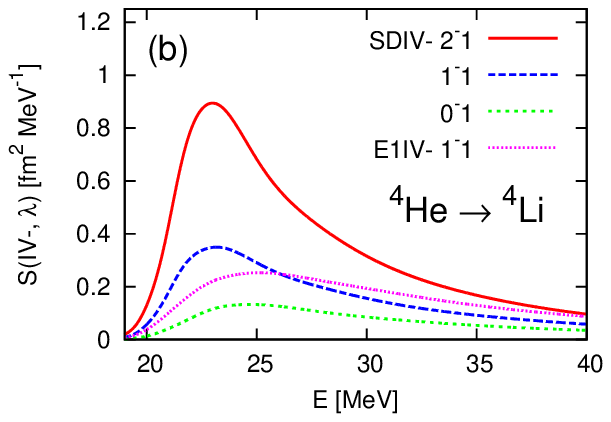,scale=1.2}
\caption{(Color online) Isovector charge-exchange SD strength
 functions of types  IV$+$ (a) and  IV$-$ (b) for
$^4$He as a function of excitation energy. 
The excitation energy is counted from the ground state of $^4$He.}
\label{SD-IVpm.fig}
\end{center}
\end{figure}

In what follows we will present results 
obtained in the Full basis calculation with the AV8$^\prime$+3NF
potential using the scaling angle $\theta=17^\circ$ unless otherwise mentioned. 
We count the excitation energy of the continuum state
from the calculated ground-state energy of $^4$He that is listed in Table 1 of 
Ref.~\cite{horiuchi12b}. 
Preliminary results on the GT and SD strength functions were
reported in Refs.~\cite{horiuchi12b,horiuchiNIC}.

Figure~\ref{SD-IV.fig} plots the SD strength 
functions of IV0 type. For 
the sake of comparison, the $E1$ strength function is also plotted by
choosing the $E1$ operator as ${\cal M}_{1\mu}=\sum_{i=1}^N{\bm \rho_i}_{\mu}
\frac{1}{2}(1-\tau_{z}(i))$. 
As seen in the figure, 
the three SD strength functions show narrower widths at their
peaks than the $E1$ strength function. Moreover their peak positions including 
the $E1$ case well correspond to the observed excitation 
energies of the four $T=1$ negative-parity states of
$^4$He~\cite{tilley92}. We will discuss this point in 
Sec.~\ref{resonance.para}.

Figure~\ref{SD-IVpm.fig} displays the charge-exchange 
SD strength functions of IV$\pm$  type 
as well as the charge-exchange $E1$ 
strength that is excited by the operator 
\begin{align}
\mathcal{M}_{1\mu}^{{\rm IV}\pm}=\sum_{i=1}^N {\bm{\rho}_i}_{\mu}
T^{{\rm IV}\pm}_i.
\end{align}
Since the mass difference between protons and neutrons is
ignored in the present calculation, we need to shift the calculated energies 
of $^4$H or $^4$Li by $\pm (m_n-m_p)$. This adjustment makes it possible 
to correctly reproduce the thresholds of $^3$H+$n$ for $^4$H and 
$^{3}$He+$p$ for $^4$Li, respectively. Similarly to the IV0 case, the
excitation energies of the charge-exchange SD peaks correspond to the
observed levels of $^4$H and $^4$Li, and their widths are narrow
compared to the charge-exchange $E1$ strength function.

\begin{figure}[th]
\begin{center}
\epsfig{file=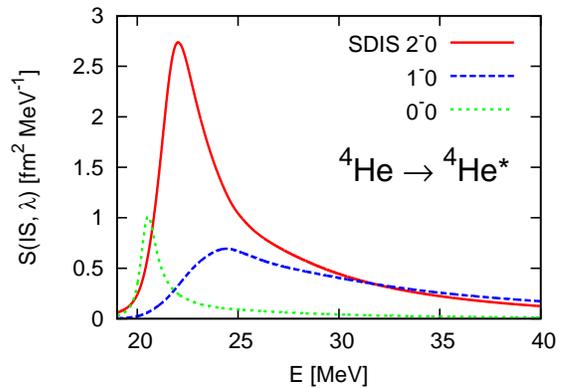,scale=1.2}
\caption{(Color online) The same as Fig.~\ref{SD-IV.fig} but for the
 isoscalar SD strength functions.}
\label{SD-IS.fig}
\end{center}
\end{figure}

We display in Fig.~\ref{SD-IS.fig} 
the IS SD strength functions that reflect 
the $J^\pi T=\lambda^-0$ continuum states of $^4$He. 
These IS SD strength functions, especially for the $0^-$ and $2^-$ cases, show 
much narrower distribution than 
the IV strength functions. These peak energies again appear to
correspond to the observed $T=0$ negative-parity levels in $^4$He.  
A close comparison between Figs.~\ref{SD-IS.fig} and \ref{SD-IV.fig} 
indicates that 
the $0^-$ case is noteworthy compared to the $1^-$ and $2^-$ cases 
in that the energy difference in the peak
positions of the same $J^-$ becomes much larger. 
As discussed in detail in Refs.~\cite{horiuchi08,horiuchi12b}, 
the reason for this is understood by analyzing the role played by the 
tensor force among others. In the previous
subsection, we mention the three negative-parity states with $T=0$ that are
observed slightly above the four-nucleon threshold and are
expected to have $d+d$ structure. Though no concentrated 
strength suggesting such states is seen in Fig.~\ref{SD-IS.fig}, 
the falloff of the IS SD strength around $28-30$ MeV 
looks flatter than 
that of the IV0 case especially in the $J^\pi=1^-$ state. This indicates  
that some IS SD strength may exist in that energy region.  To be more 
conclusive, however, a study including $d+d$ configurations 
explicitly is desirable.

\begin{figure}[ht]
\begin{center}
\epsfig{file=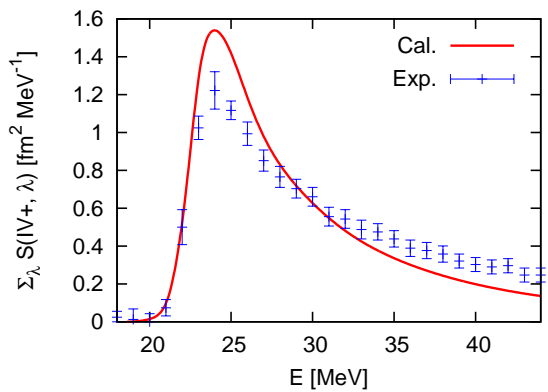,scale=1.2}
\caption{(Color online)
Summed isovector SD strength functions
 for the process 
 from the ground state of $^4$He to $^4$H as a function of 
 excitation energy. Relevant experimental data taken from
 Ref.~\cite{nakayama07} are plotted for reference. See text for the details.}
\label{SDR-exp.fig}
\end{center}
\end{figure}

To the best of our knowledge, there are no data that can directly 
be compared to the theoretical strength functions presented above. 
An only exception is 
the measurement of the charge-exchange reaction 
$^{4}$He($^7$Li,$^7$Be$\gamma$)~\cite{nakayama07,nakayama08}, from which 
the spin-nonflip ($\Delta S=0$) and spin-flip ($\Delta S=1$) components 
are separated by measuring the 0.43 MeV $\gamma$ ray of $^7$Be in
coincidence with the scattered $^7$Be. The former cross section is
ascribed to the $E1$ transition, while the latter to the SD transition. 
The shape of the deduced photoabsorption cross 
section fairly well agrees with other direct measurements using 
photons (see Fig. 9 of Ref.~\cite{horiuchi12a}), but the 
absolute magnitude is not determined definitively. 
The SD spectra corresponding to the excitation of 
the $^4$H continuum from the ground state of $^4$He is extracted from 
the spin-flip cross section in a similar way. 
Figure~\ref{SDR-exp.fig} compares
the SD strength functions of type IV$+$ 
with the `experiment'. In this figure 
the theoretical curve represents 
just a sum of the strength functions with $\lambda=0, 1, 2$, and 
the experimental distribution is normalized in such a way that both 
strength functions give the same strength when integrated 
in the energy region from $E$=18 to 44 MeV 
where the experimental data are available. 
The comparison between theory 
and experiment in Fig.~\ref{SDR-exp.fig} should thus be taken
qualitative as several assumptions are made in the analysis of 
the experiment.
The peak observed at 24 MeV agrees with the calculated one 
(see also Fig.~\ref{SD-IVpm.fig}(a)) and certainly 
it corresponds to the $J^\pi T=2^-1$ resonance of $^4$H. 
We see some difference in the shape of the strength function. 
Two conceivable reasons for it include firstly that the spin-nonflip 
process can in fact contribute to the SD 
transition as the ground state of
$^4$He contains $S=2$ components and secondly 
that some higher multipole effects 
may contribute to the cross section particularly at high 
energy~\cite{wakasa11}.  The first reason is easily understood if 
we consider the transition from $(L=S=2)$ to $(L=1, 2, 3, S=2)$. 
Further experimental information 
is needed to make a direct comparison with the calculation.

\subsection{Resonance parameters}
\label{resonance.para}

\begin{table*}
\caption{Resonance energies $E_R$ and widths $\Gamma$, given in MeV, 
of negative-parity 
levels of $A$=4 nuclei. Calculated values are extracted from 
the complex eigenvalues $E(\theta)$, 
and the SD and $E1$ strength functions $S(E)$. 
Experimental data are taken from
 Ref.~\cite{tilley92}.}
\begin{tabular}{cccccccccccccccccccccccccccccc}
\hline\hline
&&\multicolumn{7}{c}{$^4$H}&\hspace*{3mm}
&\multicolumn{7}{c}{$^4$He}&\hspace*{3mm}
&\multicolumn{7}{c}{$^4$Li}\\
\cline{3-9}\cline{11-17}\cline{19-25}
&&\multicolumn{3}{c}{$E_R$}&&\multicolumn{3}{c}{$\Gamma$}
&&\multicolumn{3}{c}{$E_R$}&&\multicolumn{3}{c}{$\Gamma$}
&&\multicolumn{3}{c}{$E_R$}&&\multicolumn{3}{c}{$\Gamma$}\\
\cline{3-5}\cline{7-9}\cline{11-13}\cline{15-17}\cline{19-21}
\cline{23-25}
$J^\pi T$&&$E(\theta)$&$S(E)$&Exp.
&&$E(\theta)$&$S(E)$&Exp.
&&$E(\theta)$&$S(E)$&Exp.
&&$E(\theta)$&$S(E)$&Exp.
&&$E(\theta)$&$S(E)$&Exp.
&&$E(\theta)$&$S(E)$&Exp.\\
\cline{1-1}\cline{3-5}\cline{7-9}\cline{11-13}\cline{15-17}\cline{19-21}
\cline{23-25}
$0^-0$  &&--&--&--&&--&--&--&&20.42&20.54 &21.01&&0.96&1.06&0.84
&&--&--&--&&--&--&--\\
$2^-0$  &&--&--&--&&--&--&--&&21.67&22.03 &21.84&&2.12&3.10&2.01
&&--&--&--&&--&--&--\\
$2^-1$  &&24.45&23.82&24.30&&5.00& 5.29 &5.42
&&23.63&23.11 &23.33&&4.99&5.58 &5.01
&&23.08 &22.99 &23.36&&5.02&6.53 &6.03\\
$1_1^-1$  &&24.68&24.04 &24.61&&5.32&6.82 &6.73
&&23.86&23.34 &23.64&&5.31&7.17 &6.20
&&23.28&23.18 &23.68&&5.36&8.06 &7.35\\
$1^-0$  &&--&--&--&&--&--&--
&&24.32&24.44 &24.25&&5.40&9.57 &6.10
&&--&--&--&&--&--&--\\
$0^-1$  &&26.51&25.46 &26.38&&7.60&9.72 &8.92
&&25.67&24.71 &25.28&&7.60&9.98 &7.97
&&25.12&24.67 &25.44&&7.69&11.03 &9.35\\
$1_2^-1$  && & 25.93 &27.13&& &12.80 &12.99
&& &25.36 &25.95&& &13.24&12.66
&& &25.15 &26.21&&   &13.92 &13.51\\
\hline\hline
\end{tabular}
\label{resSD.tab}
\end{table*}

As noted in Sec.~\ref{SDstrength.f}, all the SD and $E1$ strength functions 
exhibit some common feature: They all have one peak, though the width of
the strength distribution depends on the multipolarity $\lambda$ and 
the isospin $T$. 
It looks quite reasonable to identify the peak as a resonance. 
The resonance energy may be identified with the energy where the peak is
located. We also estimate the decay width of the resonance by 
the difference of two excitation energies at which
the strength becomes half of the maximum strength at the peak, which  
agrees with a correct width if the strength function shows the Lorentz distribution. Actually the distribution is not Lorentzian in general as we see below, but this crude estimate should be useful
as a guide. 
Table~\ref{resSD.tab} lists the resonance energies and widths of the 
negative-parity states of $^4$He, $^4$H, and $^4$Li that are determined 
in this way.
The agreement between theory and experiment is very satisfactory.  
The average deviation of the calculated resonance energies 
from experiment is 
less than 0.4 MeV for $^4$He despite the fact that most of their 
widths are larger than 5 MeV. The estimated width is also  
in reasonable agreement with experiment. 

A four-nucleon scattering calculation 
that couples $^3$H$+p$, $^3$He$+n$, and $d+d$ channels as well as 
many pseudo states is 
performed in Ref.~\cite{aoyama12} using the same Hamiltonian as the 
present study. Though the calculated phase shifts for the $J^\pi T=0^-0$ state 
show a clear resonance 
pattern at the energy consistent with the $0^-0$ level of
$^4$He, the phase shifts of the $2^-0$ and $1^-0$ states do not rise 
high enough to enable one to extract the resonance parameter. 
A more sophisticated analysis is needed to reveal resonances
using, for example, the time-delay matrix~\cite{smith60,igarashi04}.
In this context we may 
say that extracting the resonance parameter from the strength 
function is robust and can be applied to any case where even no sharp 
resonance is expected.

Since the resonance parameter obtained above is not directly 
determined from the complex eigenvalue $E_{\nu}(\theta)$ 
of the Hamiltonian, one may argue that 
the agreement is fortuitous. Of course it would be very hard to predict 
the resonance parameter correctly if a chosen operator is such that 
has only tiny strength to that resonance.  
It is therefore interesting and important to examine the complex 
eigenvalues that constitute the basis of the strength 
function. To this end we rewrite the strength function~(\ref{spectr.exp}) as
\begin{align}
S(p,\lambda,E)=\frac{1}{\pi}\sum_{\nu}
\frac{
\frac{1}{2}\gamma_{\nu}(\theta)\alpha_{\nu}^{p\lambda}(\theta)-
(E-\varepsilon_{\nu}(\theta))\beta_{\nu}^{p\lambda}(\theta)}
{(E-\varepsilon_{\nu}(\theta))^2+\frac{1}{4}(\gamma_{\nu}(\theta))^2},
\label{spectraldecomp}
\end{align}
where $\varepsilon_{\nu}(\theta)$, $\gamma_{\nu}(\theta)$, $\alpha_{\nu}(\theta)$, and $\beta_{\nu}(\theta)$ are defined by 
\begin{align}
&E_{\nu}(\theta)=\varepsilon_{\nu}(\theta)+E_0-\frac{i}{2}\gamma_{\nu}(\theta),\notag \\
&\sum_{\mu} \tilde{\mathcal{D}}^{p,\nu}_{\lambda\mu}(\theta)\mathcal{D}^{p,\nu}_{\lambda\mu}(\theta)=\alpha_{\nu}^{p\lambda}(\theta)+i\beta_{\nu}^{p\lambda}(\theta).
\end{align}
The first term of the numerator of Eq. (\ref{spectraldecomp}) gives
the Lorentz distribution, while the second term contributes 
to the background distribution.
In principle a resonance may be identified as such $E_{\nu}(\theta)$ 
that is stationary with respect to the variation of 
$\theta$~\cite{moiseyev98}. Then the strength 
function~(\ref{spectraldecomp}) has a $\theta$-independent peak around 
such a stationary energy $\varepsilon_{\nu}$. 
Resonance parameters of electron and positron complexes are in fact 
determined 
very well by examining the $\theta$-trajectory of $E_{\nu}(\theta)$~\cite{usukura02,suzuki04}. This is possible because $H(\theta)$ for the atomic case has simple structure, $H(\theta)=Te^{-2i\theta}+Ve^{-i\theta}$, where 
$T$ and $V$ are the kinetic energy and the Coulomb potential energy. 
In the nuclear case, however, $H(\theta)$ is by far complicated and a 
large-angle rotation of the nuclear potential may lead to a very long-ranged potential, 
which, together with inherent difficulties in solutions with 
the nuclear Hamiltonian, makes an accurate solution of Eq.~(\ref{ceig.eq}) extremely hard. Therefore, we first look for such eigenvalues that deviate from the rotating-continuum line as possible candidates for a resonance and 
choose the one that is closest to the peak energy of the strength function.   

For the $J^\pi T=0^-0$ and $2^-0$ states, which have
a relatively small decay width, we find only one candidate 
that may correspond to the observed resonance 
but other $J^\pi T$ states 
have two or three candidates below $4N$ threshold. 
However, no candidate is found for the $1_2^-1$ state  
that is excited by the $E1$ operator.  
The resonance energies and widths determined in this way are also listed in 
Table~\ref{resSD.tab}. The resonance energy obtained from the complex energy 
eigenvalue is in excellent agreement with experiment, even better than that determined by the strength function. The width is also satisfactorily 
reproduced. 
Two approaches to determining the resonance parameters produce 
successful results, and they are powerful, robust, and complementary.  

\begin{figure*}[th]
\begin{center}
\epsfig{file=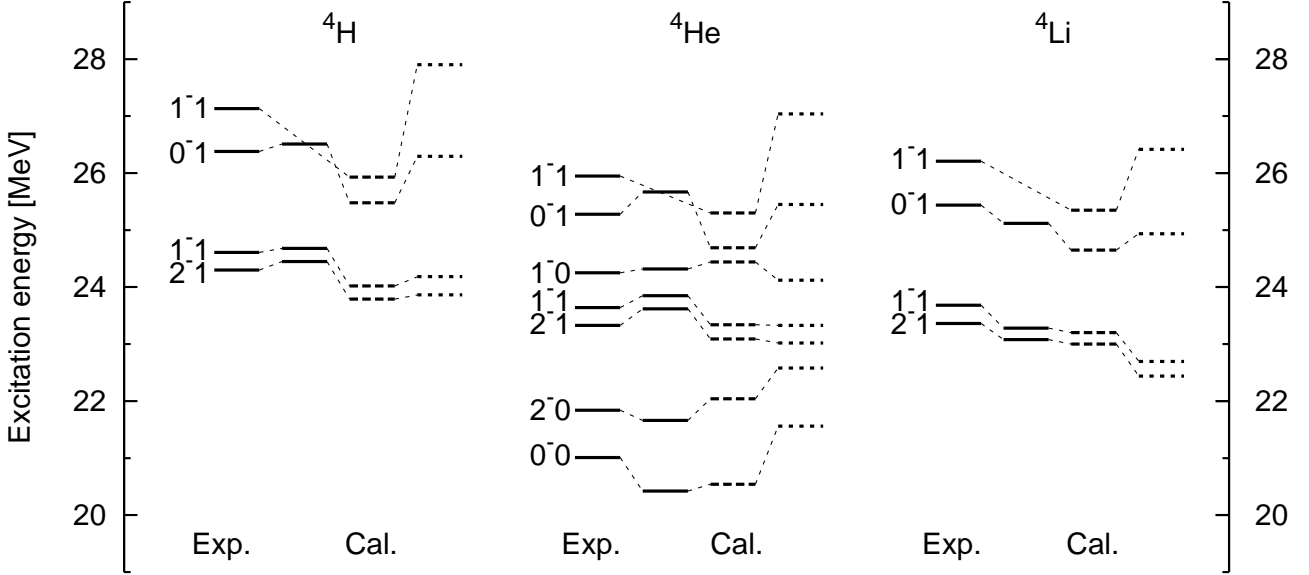,scale=1.6}
\caption{ Negative-parity levels of $A$=4
 nuclei. Excitation energies are referred to the ground-state of
 $^4$He. Solid, dashed and dotted lines of calculation are based on 
the complex eigenvalues, the SD and $E1$
strength functions, and bound-state approximation~\cite{horiuchi12b}, 
respectively. 
Experimental data are taken from Ref.~\cite{tilley92}. 
Levels belonging to the same $J^{\pi}T$ are connected by thin dotted
 lines.}
\label{spectra.fig}
\end{center}
\end{figure*}

Figure~\ref{spectra.fig} compares with experiment the resonance 
energies of the 
negative-parity states of $^4$He, $^4$H, and $^4$Li that are 
determined from the complex energy eigenvalues and the strength
functions. It is 
striking that the theory reproduces the experimental spectrum in correct 
order and moreover closely to the observed excitation energy. 
The dotted line in the figure denotes the energy obtained with a kind of 
the real stabilization method~\cite{hazi70}, that is by  
diagonalizing the Hamiltonian in the CG-GV basis 
functions~\cite{horiuchi12b}. Here the SVM search is performed 
to optimize the parameters of the basis functions by confining 
the four nucleons in some configuration space. 
It should be noted, however, that 
such calculation faces difficulty when dealing with a resonance with 
a very broad width such as the $1_2^-$ level of $^4$He, and therefore the 
resonance energy obtained in that calculation should be taken 
only approximate.

\subsection{Spin-dipole sum rules}

\label{sec.SDSR}
Sum rules are related to the energy 
moment of the strength functions  in different order and
can be expressed with the ground-state expectation
values of appropriate operators from which we 
can obtain interesting information
on the electroweak properties of nuclei
~\cite{tsuzuki84,lipparini89}.

Throughout Sec.~\ref{sec.SDSR} and Appendices A and C we denote the
numbers of nucleons, neutrons, and protons by $A$, $N$, and $Z$, 
respectively. Accordingly the center-of-mass coordinate is $\bm x_A$ 
instead of $\bm x_N$. In the other 
sections $N$ is used to denote the number of nucleons because 
the symbol $A$ is reserved to stand for the matrix that appears in 
Eq.~(\ref{GVR.eq}).

\subsubsection{Non energy-weighted sum rule}
\label{newsr}

Here we discuss the NEWSR for the SD operator
\begin{align}
m_0(p,\lambda)=\int_0^\infty S(p,\lambda,E)dE. 
\label{SD.newsr}
\end{align}
The use of the  closure relation
enables us to express the NEWSR to the expectation value of the 
operator $\sum_{\mu}\mathcal{O}_{\lambda \mu}^{p{\dagger}}
\mathcal{O}_{\lambda \mu}^p$ with respect to the ground state $\Psi_0$. 
It is convenient to express that operator 
as a scalar product of the space-space and spin-spin tensors
\begin{align}
\mathcal{Q}^p_{(\kappa)0}=\sum_{i,j=1}^A\big([\bm \rho_i \times \bm
 \rho_j]_{\kappa}\cdot [\bm \sigma_i\times\bm \sigma_j]_{\kappa}\big)
{T^p_i}^{\dagger}T^p_{j},
\label{sumopdef.eq}
\end{align}
where the rank $\kappa$ can be 0, 1, and 2, and 
the symbol $(T_{\kappa}\cdot V_{\kappa})=(-1)^{\kappa}\sqrt{2\kappa+1}[T_{\kappa}\times V_{\kappa}]_{00}$ denotes a scalar
product of spherical tensors, $T_{\kappa}$ and $V_{\kappa}$. 
As shown in Eq.~(\ref{sumdef.eq}) of Appendix A, 
the NEWSR~(\ref{SD.newsr}) is equivalently expressed, with use of 
$U_{\lambda \kappa}$ of Eqs.~(\ref{umatrix}) and (\ref{ucoef.eq}), as 
\begin{align}
m_0(p,\lambda)=\sum_{\kappa=0}^2 U_{\lambda \kappa} 
\langle \mathcal{Q}^p_{(\kappa)0} \rangle.
\label{derivesum1.eq}
\end{align}
The expectation value, 
$\langle \mathcal{Q}^p_{(\kappa)0} \rangle=\langle \Psi_0 |\mathcal{Q}^p_{(\kappa)0}|\Psi_0 \rangle$, 
can be evaluated using the basis 
functions~(\ref{lscoupled}), (\ref{spin.fn}), and (\ref{GVR.eq}), as
explained in Appendix B. 

In order to check the extent to which the NEWSR is satisfied, 
we compare $m_0(p,\lambda)$ that is calculated separately with 
Eq.~(\ref{SD.newsr}) or with Eq.~(\ref{derivesum1.eq}). 
Table~\ref{m-values.tab} lists the calculated NEWSR for the SD strength 
functions. We also list the values of 
$\langle \mathcal{Q}^p_{(\kappa)0} \rangle$ in Table~\ref{Q-values.tab} 
for the sake of discussions below. As seen in Table~\ref{m-values.tab}, 
the two different ways
of calculating the sum rules give virtually the same result 
for both cases of AV8$^\prime$+3NF and G3RS+3NF interactions, which 
is never trivial because we use the fully correlated 
ground-state wave function for $^4$He. The perfect agreement
confirms that the basis 
functions prepared for the description of the SD excitation are 
sufficient enough to account for all the strength in the continuum. 
The NEWSR calculated with Eq.~(\ref{derivesum1.eq}) for 
the Minnesota (MN) potential~\cite{MN} is also listed in 
Table~\ref{m-values.tab}. 
A comparison of the central MN force case with the 
realistic potentials will be useful to know how much the 
sum rule is affected by the tensor force.

\begin{table*}
\caption{Non energy-weighted sum rules of the SD strength 
functions, in units of fm$^2$, calculated from different models 
for the nucleon-nucleon potentials. The sum rules 
calculated by Eqs.~(\ref{SD.newsr}) and (\ref{derivesum1.eq}) are 
labeled $m_0(p,\lambda)$ and SR, respectively.}
\begin{tabular}{ccccccccccccccccccccccccc}
\hline\hline
&&\multicolumn{8}{c}{AV8$^\prime$+3NF}&\hspace*{3mm}
&\multicolumn{8}{c}{G3RS+3NF}&\hspace*{3mm}
&\multicolumn{5}{c}{MN}\\
\cline{3-10}\cline{12-19}\cline{21-25}
&&\multicolumn{2}{c}{IS}&&\multicolumn{2}{c}{IV0}&&\multicolumn{2}{c}{IV$\pm$}
&&\multicolumn{2}{c}{IS}&&\multicolumn{2}{c}{IV0}&&\multicolumn{2}{c}{IV$\pm$}
&&IS&&IV0&&IV$\pm$\\
\cline{3-4}\cline{6-7}\cline{9-10}\cline{12-13}\cline{15-16}\cline{18-19}
\cline{21-21}\cline{23-23}\cline{25-25}
$\lambda$
&&$m_0(p,\lambda)$&SR&&$m_0(p,\lambda)$&SR&&$m_0(p,\lambda)$&SR
&&$m_0(p,\lambda)$&SR&&$m_0(p,\lambda)$&SR&&$m_0(p,\lambda)$&SR
&&SR&&SR&&SR\\
\cline{1-1}\cline{3-4}\cline{6-7}\cline{9-10}\cline{12-13}\cline{15-16}\cline{18-19}
\cline{21-21}\cline{23-23}\cline{25-25}
0&&2.71 &2.71 &&4.59 &4.59&&2.30&2.30
&&2.83 &2.84&&4.74 &4.74&&2.37&2.37&&3.90&&3.49&&1.74\\
1&&12.16 &12.17 &&9.35 &9.36&&4.68&4.68
&&12.64 &12.65&&9.72 &9.73&& 4.86&4.86&&11.71&&10.46&&5.23\\
2&&17.98 &18.02 &&18.36 &18.38&&9.18&9.19
&&18.77 &18.79&&19.02 &19.04&&9.51&9.52&&19.51&&17.43&&8.71\\
\hline\hline
\end{tabular}
\label{m-values.tab}
\end{table*}

\begin{table*}
\caption{Expectation values of $\mathcal{Q}_{(\kappa)0}^p$ and 
its one- and two-body terms with respect to the ground state of $^4$He. 
Values are given in units of fm$^2$.}
\begin{tabular}{lccccccccccccc}
\hline\hline
&&\multicolumn{3}{c}{AV8$^\prime$+3NF}&\hspace*{3mm}
&\multicolumn{3}{c}{G3RS+3NF}&\hspace*{3mm}
&\multicolumn{3}{c}{MN}\\
\cline{3-5}\cline{7-9}\cline{11-13}
 &&IS&IV0&IV$\pm$&&IS&IV0&IV$\pm$&&IS&IV0&IV$\pm$\\
\cline{3-5}\cline{7-9}\cline{11-13}
$\langle \mathcal{Q}_{(0)0}^{p}\rangle$&&10.97  &10.78&5.39
&&11.42&11.17&5.59&&11.71&10.46&5.23\\
\cline{1-1}\cline{3-5}\cline{7-9}\cline{11-13}
$\langle\mathcal{Q}_{(0)0}^{p(1)}
\rangle$&&8.41&8.41&4.21&&8.66&8.66&4.33&&7.96&7.96&3.98\\
$\langle\mathcal{Q}_{(0)0}^{p(2)}\rangle$
&&2.56&2.37&1.18&&2.76&2.51&1.25&&3.75&2.50&1.25\\
\hline
$\langle\mathcal{Q}_{(1)0}^{p}\rangle$
&&0.21&$-$0.08&$-$0.04&&0.24&$-$0.09&$-$0.04&&0.00&0.00&0.00\\
\cline{1-1}\cline{3-5}\cline{7-9}\cline{11-13}
$\langle\mathcal{Q}_{(1)0}^{p(1)}\rangle$&&--&--&--&&--&--&--&&--&--&--\\
$\langle\mathcal{Q}_{(1)0}^{p(2)}\rangle$
&&0.21&$-$0.08&$-$0.04&&0.24&$-$0.09&$-$0.04&&0.00&0.00&0.00\\
\hline
$\langle\mathcal{Q}_{(2)0}^{p}\rangle$
&&$-$2.61&2.92&1.46&&$-$2.68&2.97&1.49&&0.00&0.00&0.00\\
\cline{1-1}\cline{3-5}\cline{7-9}\cline{11-13}
$\langle\mathcal{Q}_{(2)0}^{p(1)}\rangle$&&--&--&--&&--&--&--&&--&--&--\\
$\langle\mathcal{Q}_{(2)0}^{p(2)}\rangle$
&&$-$2.61&2.92&1.46&&$-$2.68&2.97&1.49&&0.00&0.00&0.00\\
\hline\hline
\end{tabular}
\label{Q-values.tab}
\end{table*}

Among the three expectation values of $\langle \mathcal{Q}^p_{(\kappa)0}
\rangle$ in Eq.~(\ref{derivesum1.eq}), 
the $\kappa=0$ term gives a dominant contribution to the NEWSR. 
See Table~\ref{Q-values.tab}. This is obviously because the 
major component of the ground state of $^4$He is $S=0$ and it 
has a non-vanishing expectation value only for  $\mathcal{Q}^p_{(0)0}$. 
In this limiting case  
$m_0(p, \lambda)$  is proportional to $U_{\lambda 0}$. 
Therefore the $\lambda$-dependence of the NEWSR turns out 
to be $1:3:5$ for $\lambda=0,\, 1,\, 2$, independently of $p$. 
This rule is confirmed in the MN case of  
Table~\ref{m-values.tab}. The deviation from this ratio is due 
to the contributions of other $\mathcal{Q}^p_{(\kappa)0}$ terms, 
especially the $\kappa=2$ term. The $\mathcal{Q}^p_{(2)0}$ term 
contributes to the NEWSR through the coupling matrix element 
between the $S=0$ and $S=2$ components of the ground state of
$^4$He. Since the 
admixture of the $S=2$ component is primarily determined by the 
tensor force, the deviation reflects the tensor correlations 
in the ground state. 
Neglecting the minor contribution 
of $\mathcal{Q}^p_{(1)0}$, Eq.~(\ref{derivesum1.eq}) 
suggests that $m_0(p, \lambda)$ is very well approximated by 
\begin{align}
m_0(p, 0)& =\frac{1}{3}(\langle \mathcal{Q}^p_{(0)0} \rangle -
\langle \mathcal{Q}^p_{(1)0} \rangle 
+ \langle \mathcal{Q}^p_{(2)0} \rangle),  \notag \\
m_0(p, 1)&=m_0(p, 0)+\frac{1}{2}(\langle \mathcal{Q}^p_{(1)0} \rangle 
-3 \langle \mathcal{Q}^p_{(2)0} \rangle)\notag \\
&\approx m_0(p, 0)  -\frac{3}{2}\langle \mathcal{Q}^p_{(2)0} \rangle,
\notag \\
m_0(p, 2)&=\frac{5}{3}m_0(p, 0)+\frac{1}{2}(5\langle \mathcal{Q}^p_{(1)0} \rangle 
-3 \langle \mathcal{Q}^p_{(2)0} \rangle)\notag \\
&\approx \frac{5}{3}m_0(p, 0)-\frac{3}{2}\langle \mathcal{Q}^p_{(2)0} \rangle.
\end{align}
Thus the deviation of the ratio from $1:3:5$ is simply controlled 
by $-\frac{3}{2}\langle \mathcal{Q}^p_{(2)0} \rangle$, which is  
very well satisfied in the examples of Table~\ref{m-values.tab}. 
Since $\langle \mathcal{Q}^{p}_{(2)0} \rangle$ 
is negative for $p={\rm IS}$, the ratio further increases from $1:3:5$, whereas 
it is positive for $p={\rm IV0}$ and IV$\pm$, and 
the ratio approximately reduces to $1:2:4$.

As discussed above, $\langle \mathcal{Q}^p_{(\kappa)0} \rangle$ plays 
a central role to determine the NEWSR for the SD strength 
functions. Inverting Eq.~(\ref{derivesum1.eq}) 
makes it possible to express $\langle \mathcal{Q}^p_{(\kappa)0} \rangle$
as a sum, over the multipole $\lambda$, of the NEWSR 
\begin{align}
\langle \mathcal{Q}^p_{(\kappa)0} \rangle=
\sum_{\lambda=0}^2U^{-1}_{\ \ \kappa \lambda}m_0(p, \lambda),
\label{mtoQ}
\end{align}
where $U^{-1}$ is the inverse matrix of $U$ as given in Eq.~(\ref{uinv}). 
If the NEWSR for all $\lambda$ are experimentally measured, 
the above equation indicates that 
$\langle \mathcal{Q}^p_{(\kappa)0} \rangle$ for all 
$\kappa$ can be determined  from experiment. 
Some examples are 
\begin{align}
&3\langle \mathcal{Q}^p_{(0)0} \rangle=m_0(p, 0)+m_0(p, 1)+ m_0(p, 2),
\notag \\
&6\langle \mathcal{Q}^p_{(2)0} \rangle=10 m_0(p, 0)-5 m_0(p, 1)+ m_0(p, 2).
\end{align}

To clarify the physical meaning of the operator $\mathcal{Q}^p_{(\kappa)0}$, 
it is instructive to decompose it into one- and two-body terms: 
\begin{align}
\mathcal{Q}^p_{(\kappa)0}=
\mathcal{Q}^{p(1)}_{(\kappa)0}+\mathcal{Q}^{p(2)}_{(\kappa)0},
\end{align}
where
\begin{align}
&\mathcal{Q}^{p(1)}_{(\kappa)0}=\delta_{\kappa 0}\sum_{i=1}^A\bm
 \rho_i^2{T^p_i}^{\dagger}T^p_{i},\notag \\
&\mathcal{Q}^{p(2)}_{(\kappa)0}=\sum_{j>i=1}^A\big([\bm \rho_i \times \bm
 \rho_j]_{\kappa}\cdot [\bm \sigma_i\times\bm \sigma_j]_{\kappa}\big)T^p_{ij}
\label{qtb.eq}
\end{align}
with
\begin{align}
T^p_{ij}= {T^p_i}^{\dagger}T^p_{j}+{T^p_j}^{\dagger}T^p_{i}.
\end{align}
The isospin operators in Eq.~(\ref{qtb.eq}) are simplified with use of
Eq.~(A2): ${T^p_i}^{\dagger}T^p_{i}$ is 1 for $p={\rm IS}$, IV0,
and $(1\mp \tau_z(i))/2$ for $p={\rm IV}\pm$, whereas  $T^p_{ij}$ is 
2 for $p={\rm IS}$, $2\tau_z(i)\tau_z(j)$ for $p={\rm IV0}$, and 
$((\bm \tau(i)\cdot \bm \tau(j))-\tau_z(i)\tau_z(j))/2$ 
for $p={\rm IV}\pm$, respectively.  
The one-body term is spin-independent and 
appears only for $\kappa=0$, which gives the largest contribution
to the NEWSR. The two-body term with $\kappa=2$ is 
particularly interesting
because it contains the tensor operator characteristic of 
the one-pion-exchange potential. See Appendix A for detail.

The expectation value 
of the one-body term is expressed in terms of the 
root-mean-square radius of nucleon distribution in the ground state 
\begin{align}
&\langle \mathcal{Q}^{{\rm IS}(1)}_{(0)0} \rangle=\langle \mathcal{Q}^{{\rm IV0}(1)}_{(0)0} \rangle=A \langle r_N^2 \rangle, \notag \\ 
&\langle \mathcal{Q}^{{\rm IV+}(1)}_{(0)0} \rangle=Z\langle r_p^2 \rangle, \quad \langle \mathcal{Q}^{{\rm IV-}(1)}_{(0)0} \rangle=N\langle r_n^2 \rangle.
\end{align}
Noting that the two-body term $\mathcal{Q}^{{\rm IV}+(2)}_{(\kappa)0}$
is identical to $\mathcal{Q}^{{\rm IV}-(2)}_{(\kappa)0}$ for any
$\kappa$, we obtain the following well-known relation between the
NEWSR~\cite{gaarde81} 
\begin{align}
m_0({\rm IV}-, \lambda)- m_0({\rm IV}+, \lambda)
=\frac{2\lambda +1}{3}
\big(N\langle r_n^2 \rangle - Z\langle r_p^2 \rangle \big).
\end{align}
This difference vanishes in the present case because the isospin
impurity of the ground-state of $^4$He is ignored.

\subsubsection{Energy-weighted sum rule}

Now we discuss the EWSR for the SD operator.
The SD EWSR can be derived in the same manner as the $E1$ operator, 
and it is expressed as
\begin{align}
m_1(p,\lambda)&=\int_0^\infty ES(p,\lambda,E)dE\notag\\
&=\langle X^p_{(\lambda)0}(H) \rangle,
\label{ewsr.def}
\end{align}
where $X^p_{(\lambda)0}(H)$ denotes the double commutator
of the Hamiltonian with the SD operator 
\begin{align}
X^p_{(\lambda)0}(H)=\frac{1}{2}\sum_\mu [\mathcal{O}_{\lambda\mu}^{p\dagger},
[H,\mathcal{O}_{\lambda\mu}^p]].
\label{DBC.eq}
\end{align}
The double commutator of the kinetic energy operator 
$T=\sum_{i=1}^AT_i-T_{\rm cm}$ is worked 
out in Appendix C. The commutator was considered in 
Ref.~\cite{tsuzuki79} for IS and IV0 cases.  
The result for all SD cases is summarized as 
\begin{align}
&X^{p}_{(\lambda)0}(T)=\frac{(A-1)\hbar^2}{2Am_N}(2\lambda+1)N^p\notag \\
& -\frac{\hbar^2}{6Am_N}(2\lambda+1)
\sum_{j>i=1}^A(\bm \sigma_i\cdot\bm \sigma_j)T^p_{ij}\notag \\ 
& -\frac{i\hbar}{6m_N} (2\lambda+1)\sum_{i=1}^A
\big(\bm \rho_i\cdot(\bm p_i-\textstyle{\frac{1}{A}}\bm P_{\rm tot})\big)[{T^p_{i}}^{\dagger}, T^p_i]\notag \\ 
&+\frac{\hbar}{6m_N}C_{\lambda}^p\sum_{i=1}^A
\big((\bm \rho_i \times (\bm p_i-\textstyle{\frac{1}{A}}\bm P_{\rm tot}))
 \cdot \bm \sigma_i\big),
\label{sd.ewsr.formula}
\end{align}
where $\bm P_{\rm tot}=\sum_{i=1}^A{\bm p}_i$ is the total momentum and 
$C^p_{\lambda}$ is related to $C_{\lambda}$ of Eq.~(C7) as 
\begin{align}
C_{\lambda}^{\rm IS}=C_{\lambda}^{\rm IV0}=C_{\lambda},\quad 
C_{\lambda}^{{\rm IV}+}=C_{\lambda}^{{\rm IV}-}=\frac{1}{2}C_{\lambda},
\end{align}
and $N^p=\sum_{i=1}^AT_i^p{T^p_i}^{\dagger}$ reduces to $A$ for $p={\rm IS}$, 
IV0, $A-Z$ for $p={\rm IV}+$, and $A-N$ for $p={\rm IV}-$, respectively. 
The isospin commutator $[{T^p_{i}}^{\dagger}, T^p_i]$ 
vanishes for $p={\rm IS}$ and IV0, while it reduces to 
$\mp \tau_z(i)$ for $p={\rm IV}\pm$.
The round bracket $(\bm a \times \bm b)$ stands for
the vector product of $\bm a$ and $\bm b$, 
$(\bm a \times \bm b)_{\mu}=-\sqrt{2}i[\bm a \times \bm b]_{1\mu}$. 

We name the four terms on the right-hand side of 
Eq.~(\ref{sd.ewsr.formula}) as model-independent (MI), spin-spin (SS), 
dilation (DL), and spin-orbit (SO) terms, respectively. The name of 
dilation is adopted because $\big(\bm \rho_i\cdot(\bm
p_i-\textstyle{\frac{1}{A}}\bm P_{\rm tot})\big)$ is a generator for the
dilation operator. The MI term 
makes a contribution to the SD EWSR, independently of the ground-state 
wave function. Thus the kinetic energy contribution to the EWSR 
becomes model-independent in so far as the contribution of the other 
terms can be neglected compared to the MI term. 
For a fixed $p$ the $\lambda$-dependence of each term  
is simply given by $2\lambda+1$
except for the SO term, which changes according to the ratio 
of $2:3:(-5)$ for $\lambda=0,1,2$. On the other hand, 
for a fixed $\lambda$ the $p$-dependence 
of the four terms is a little complicated. The MI
term changes in proportion to $A:A:A-Z:A-N$, while the SO term 
is in ratio of $1:1:1/2:1/2$ for $p={\rm IS}$, IV0, IV$+$, 
IV$-$, respectively. The DL term identically vanishes for $p={\rm IS}$ and
IV0, and furthermore it turns out to have no contribution to the 
EWSR even for $p={\rm IV}\pm$ because no isospin mixing is taken into account 
in our ground state of $^4$He. 

Table~\ref{EWSR.tab} 
lists the values of $m_1(p,\lambda)$ together with
the contributions of the kinetic energy term and its 
four terms to the EWSR calculated using the AV8$^\prime$+3NF 
and G3RS+3NF potentials. 
The EWSR slightly depends on the potential
models particularly for the IS SD strengths.
Even in those cases the contribution of the kinetic energy 
to the EWSR remains almost the same. The contribution
of the MI term to $\langle X_{(\lambda)0}^p(T)\rangle$ is found to be 
more than 74 \% for all the cases, and really occupies a main portion of the 
kinetic energy contribution. 
The two interactions give almost the same contribution for the SS terms.
Though the SO terms show some dependence on the interactions,
the kinetic energy contributions $\langle X^p_{(\lambda)0}(T)\rangle$
are found to be approximately model-independent.

\begin{table*}
\caption{Energy-weighted sum rules of the SD strength functions,
$m_1(p,\lambda)$, in units of fm$^2$MeV,
calculated from different models 
for the nucleon-nucleon potentials. Contribution of each term 
of the kinetic energy to the sum rule is also listed. 
See text for the details. }
\begin{tabular}{ccccccccccccccccccc}
\hline\hline
&&\multicolumn{15}{c}{AV8$^\prime$+3NF}\\
\cline{3-17}
   &&\multicolumn{3}{c}{IS}&\hspace*{2mm}
&\multicolumn{3}{c}{IV0}&\hspace*{2mm}
&\multicolumn{3}{c}{IV$+$}&\hspace*{2mm}
&\multicolumn{3}{c}{IV$-$}\\
\cline{3-5}\cline{7-9}\cline{11-13}\cline{15-17}
&& $\lambda=0$  &$\lambda=1$  &$\lambda=2$
&& $\lambda=0$  &$\lambda=1$  &$\lambda=2$
&& $\lambda=0$  &$\lambda=1$  &$\lambda=2$
&& $\lambda=0$  &$\lambda=1$  &$\lambda=2$\\
\cline{3-5}\cline{7-9}\cline{11-13}\cline{15-17}
$m_1(p,\lambda)$&&126&782&949&&218&450&766&&110&227&389&&109&225&383\\
\cline{1-1}\cline{3-5}\cline{7-9}\cline{11-13}\cline{15-17}
$\langle X_{(\lambda)0}^p(T)\rangle$&&74.4&227&392&&78.2&239&411
&&39.1&119&205&&39.1&119&205\\
\cline{1-1}\cline{3-5}\cline{7-9}\cline{11-13}\cline{15-17}
MI&&62.2&187&311&&62.2&187&311&&31.1&93.3&156&&31.1&93.3&156\\
SS&&14.9&44.6&74.3&&18.7&55.9&93.2&&9.32&27.9&46.6&&9.32&27.9&46.6\\
DL&&--&--&--&&--&--&--&&0.00&0.00&0.00&&0.00&0.00&0.00\\
SO&&$-$2.62&$-$3.92&6.54&&$-$2.62&$-$3.92&6.54&&$-$1.31&$-$1.96&3.27&&$-$1.31&$-$1.96&3.27\\
\cline{1-1}\cline{3-5}\cline{7-9}\cline{11-13}\cline{15-17}
$m_1(p,\lambda)-\langle X_{(\lambda)0}^p(T)\rangle$
&&51.5&555&557&&139&211&355&&71.3&108&183&&70.1&106&178\\
\hline
&&\multicolumn{15}{c}{G3RS+3NF}\\
\cline{3-17}
   &&\multicolumn{3}{c}{IS}&&\multicolumn{3}{c}{IV0}
   &&\multicolumn{3}{c}{IV$+$}&&\multicolumn{3}{c}{IV$-$}\\
\cline{3-5}\cline{7-9}\cline{11-13}\cline{15-17}
&& $\lambda=0$  &$\lambda=1$  &$\lambda=2$
&& $\lambda=0$  &$\lambda=1$  &$\lambda=2$
&& $\lambda=0$  &$\lambda=1$  &$\lambda=2$
&& $\lambda=0$  &$\lambda=1$  &$\lambda=2$\\
\cline{3-5}\cline{7-9}\cline{11-13}\cline{15-17}
$m_1(p,\lambda)$&&111&697&843&&202&426&723&&104&216&370&&102&213&363\\
\cline{1-1}\cline{3-5}\cline{7-9}\cline{11-13}\cline{15-17}
$\langle X_{(\lambda)0}^p(T)\rangle$&&73.2&227&403&&76.3&236&419
&&38.1&118&209&&38.1&118&209\\
\cline{1-1}\cline{3-5}\cline{7-9}\cline{11-13}\cline{15-17}
MI&&62.2&187&311&&62.2&187&311&&31.1&93.3&156&&31.1&93.3&156\\
SS&&16.0&47.9&79.8&&19.0&57.1&95.1&&9.51&28.5&47.6&&9.51&28.5&47.6\\
DL&&--&--&--&&--&--&--&&0.00&0.00&0.00&&0.00&0.00&0.00\\
SO&&$-$4.97&$-$7.45&12.4&&$-$4.97&$-$7.45&12.4&&$-$2.48&$-$3.73&6.21&&$-$2.48&$-$3.73&6.21\\
\cline{1-1}\cline{3-5}\cline{7-9}\cline{11-13}\cline{15-17}
$m_1(p,\lambda)-\langle X_{(\lambda)0}^p(T)\rangle$
&&37.8&470&439&&126&189&304&&65.5&97.8&160&&63.5&94.6&153\\
\hline\hline
\end{tabular}
\label{EWSR.tab}
\end{table*}

The enhancement of the computed sum
rule~(\ref{ewsr.def}) compared to $\langle X_{(\lambda)0}^p(T)\rangle$
indicates the contribution of the potential energy to the EWSR. The 
enhancement factor for the $E1$ operator is $1.0-1.1$ for the present 
nuclear forces~\cite{horiuchi12a}. 
The AV8$^\prime$ potential has a stronger tensor component 
than the G3RS potential. Because of this the tensor potential 
($S_{ij}\bm \tau_i\cdot \bm \tau_j$) of the AV8$^\prime$ potential 
gives the larger contribution to the $E1$ EWSR.  
In the SD case, however, the 
enhancement is more complicated and depends on 
both multipolarity $\lambda$ and isospin label $p$. 
To elucidate this further, we have to calculate the double commutator for each 
piece of the nucleon-nucleon potential as in the kinetic 
energy and evaluate its ground-state expectation value.

\section{Conclusions}

\label{conclusions.sec}

We study both isovector and isoscalar spin-dipole (SD) strength 
functions in four-body calculations using realistic
nuclear forces. Two different potentials are employed 
to see the sensitivity on the $D$-state probability 
produced by the tensor correlation. 
The SD excitation is built on the ground state of $^4$He that 
is described accurately with use of explicitly correlated 
Gaussian bases. 
The continuum states including two- and three-body 
decay channels are discretized in the correlated Gaussians 
with aid of the complex scaling method.

Experimental data that can directly be compared to the 
calculation are presently only the resonance parameters of the 
negative-parity levels of $A=4$ nuclei. Both the resonance 
energies and widths deduced from the SD and electric-dipole 
strength functions or the eigenvalues of the complex-scaled 
Hamiltonian are all in fair agreement with experiment. 
This success is never trivial considering that most of the resonances 
among 15 levels have broad widths larger than 5 MeV. A combined use 
of both complex energies and appropriate strength functions provides 
us with a robust tool to determine resonance parameters. 

The non energy-weighted sum rule (NEWSR) of the SD strength function 
is investigated by relating it to 
the expectation values of three scalar products of the space-space and 
spin-spin tensors with respect to the ground state of $^4$He. 
It turns out that our model space satisfies 
the NEWSR for each SD operator perfectly.  
The tensor operator of rank 2, $\mathcal{Q}^p_{(2)0}$, 
is sensitive to the $D$-state correlation 
in the ground state induced by the tensor force, and it is mainly 
responsible for distorting the ratio of the NEWSRs for the 
multipolarity $\lambda=0, 1, 2$ from the uncorrelated ratio of $1:3:5$. 
An experimental observation of this ratio is desirable since it 
may lead us to reveal the degree of tensor correlations in 
the ground state. The energy-weighted sum rule (EWSR) for the SD operator 
is also examined. A formula is derived to calculate the contribution of 
the kinetic energy to the EWSR. The difference between the 
EWSR and the kinetic energy contribution shows some dependence on 
$\lambda$ as wells as the isospin character of the SD operator. 
Further study is needed to clarify the origin of its dependence by analyzing 
the contribution of each piece of the nuclear potential. 

Other $T=0$ resonances with $0^-,\,1^-,\, 2^-$ and $1^+,\, 2^+$ 
exist in $^4$He above and below the $2n+2p$ threshold. It would be 
interesting to investigate these levels by the isoscalar 
SD excitation and some appropriate excitations produced by
e.g., isoscalar quadrupole, magnetic dipole, and spin-quadrupole operators
with further attention being paid to $d+d$ type configurations. 
 
The SD strength functions are important inputs for evaluating 
neutrino-nucleus reaction rates. 
A calculation of neutrino-$^4$He reaction rate is in progress as a 
consequence of the present study. It is desirable that the predicted SD
strength functions are tested with experimental measurements 
in order for such reaction rate calculation to be precise. 

\section*{Acknowledgments}

The authors thank T. Sato for valuable discussions 
on the electroweak processes and S. Nakayama for useful communications on the
SD experimental data of $^4$He. The work of Y.~S. is supported
in part by Grants-in-Aid for Scientific Research 
(No. 21540261 and No. 24540261) of
the Japan Society for the Promotion of Science.

\appendix

\section{Multipole decomposition of the spin-dipole non
energy-weighted sum rule}
\label{app.multipole}

Here we derive Eqs.~(\ref{sumopdef.eq}) and (\ref{derivesum1.eq}) 
by decomposing the operator 
$\sum_{\mu}\mathcal{O}_{\lambda\mu}^{p\dagger}\mathcal{O}_{\lambda
\mu}^p$ into multipoles. 
Substituting Eq.~(\ref{sd.op}) in
$\sum_{\mu}\mathcal{O}_{\lambda\mu}^{p\dagger}\mathcal{O}_{\lambda
\mu}^p$ and recoupling the coordinate and spin operators, we obtain 
\begin{align}
&\sum_{\mu}\mathcal{O}_{\lambda\mu}^{p\dagger}\mathcal{O}_{\lambda \mu}^p
 \notag \\
&=(-1)^{\lambda}\sum_{i,j=1}^A\big([\bm \rho_i \times \bm
 \sigma_i]_{\lambda}\cdot [\bm \rho_j \times \bm
 \sigma_j]_{\lambda}\big){T^p_i}^{\dagger}T^p_{j}
\notag \\
&=\sum_{\kappa}U_{\lambda \kappa}\mathcal{Q}^p_{(\kappa)0}.
\label{sumdef.eq}
\end{align}
The isospin operator 
${T^p_i}^{\dagger}T^p_{j}$ reads 
\begin{align}
1,\quad  \tau_{z}(i)\tau_{z}(j), \quad (\bm t_i \cdot \bm
 t_j)-t_{z}(i)t_{z}(j)\pm i(\bm t_i\times \bm t_j)_z
\end{align}
for $p={\rm IS}$, IV0, and IV$\pm$, respectively. 
The coefficient $U_{\lambda \kappa}$ is expressed by unitary 
Racah coefficients $U$ as 
\begin{align}
U_{\lambda\kappa}=(-1)^{\lambda}\sqrt{\frac{2\lambda+1}{2\kappa+1}}U(1111,\lambda\kappa), 
\label{umatrix}
\end{align}
or more explicitly 
\begin{align}
(U_{\lambda \kappa})=
\begin{pmatrix}
\frac{1}{3}&-\frac{1}{3}&\frac{1}{3}\\
1&-\frac{1}{2}&-\frac{1}{2}\\
\frac{5}{3}&\frac{5}{6}&\frac{1}{6}
\end{pmatrix}, 
\label{ucoef.eq}
\end{align}
where both row and column labels, $\lambda$ and $\kappa$, are arranged
in order of 0, 1, and 2. The inverse of the matrix $(U_{\lambda
\kappa})$, 
\begin{align}
({U^{-1}}_{\kappa \lambda})=
\begin{pmatrix}
\frac{1}{3}&\frac{1}{3}&\frac{1}{3}\\
-1&-\frac{1}{2}&\frac{1}{2}\\
\frac{5}{3}&-\frac{5}{6}&\frac{1}{6}\\
\end{pmatrix},
\label{uinv} 
\end{align}
is used to obtain the expectation value of 
$\mathcal{Q}^p_{(\kappa)0}$ with respect to the ground state as discussed
in Sec.~\ref{newsr}. See Eq.~(\ref{mtoQ}).

The multipole operator $\mathcal{Q}^p_{(\kappa)0}$ consists of one- and
two-body terms
\begin{align}
\mathcal{Q}^p_{(\kappa)0}=\mathcal{Q}^{p(1)}_{(\kappa)0}+\mathcal{Q}^{p(2)}_{(\kappa)0}
\end{align}
as shown in Eq.~(\ref{qtb.eq}). 
The two-body term with $\kappa=2$ is of particular interest because 
it contains the tensor operator. To see this, it is convenient to 
rewrite $\mathcal{Q}^{p(2)}_{(\kappa)0}$ in terms of the relative
and center-of-mass coordinates of two nucleons rather than the 
single-particle like coordinates, $\bm \rho_i$ and $\bm \rho_j$. 
By introducing the 
coordinates $\bm r_{ij}$ and $\bm R_{ij}$ by 
\begin{align}
\bm{r}_{ij}&=\bm \rho_i-\bm \rho_j=\bm{r}_i-\bm{r}_j,\notag \\
\bm{R}_{ij}&=\frac{1}{2}(\bm \rho_i+\bm \rho_j)=\frac{1}{2}(\bm{r}_i+\bm{r}_j)-\bm{x}_A,
\end{align}
$\mathcal{Q}^{p(2)}_{(\kappa)0}$ is decomposed to three terms:
\begin{align}
\mathcal{Q}^{p(2)}_{(\kappa)0}=\mathcal{Q}^{p(2)r}_{(\kappa)0}+
\mathcal{Q}^{p(2)R}_{(\kappa)0}+
\delta_{\kappa 1}\mathcal{Q}^{p(2)rR}_{(1)0}, 
\label{opdecom.eq}
\end{align}
where
\begin{align}
&\mathcal{Q}^{p(2)r}_{(\kappa)0}=-\frac{1}{4}
\sum_{j>i=1}^A\big([\bm
 r_{ij}\times \bm r_{ij}]_{\kappa}\cdot [\bm \sigma_i\times \bm
 \sigma_j]_{\kappa}\big)T^p_{ij}, \notag \\
&\mathcal{Q}^{p(2)R}_{(\kappa)0}=\sum_{j>i=1}^A\big([\bm
 R_{ij}\times \bm R_{ij}]_{\kappa}\cdot [\bm \sigma_i\times \bm
 \sigma_j]_{\kappa}\big)T^p_{ij}, \notag \\
&\mathcal{Q}^{p(2)rR}_{(1)0}=\sum_{j>i=1}^A
\big([\bm r_{ij}\times \bm R_{ij}]_{1}\cdot [\bm \sigma_i\times \bm
 \sigma_j]_{1}\big)T^p_{ij}.
\label{qtbdecom.eq}
\end{align}
The operators $\mathcal{Q}^{p(2)r}_{(\kappa)0}$ and 
$\mathcal{Q}^{p(2)R}_{(\kappa)0}$ have non-vanishing contributions only 
for $\kappa=0$ and 2. It is easy to see that 
the $\mathcal{Q}^{p(2)r}_{(2)0}$ term contains the tensor operator $S_{ij}$. 

\section{Calculation of the matrix elements of 
quadratic spatial tensors}

In this appendix, we give a formula of calculating
the matrix element of $\mathcal{Q}^p_{(\kappa)0}$, Eq.~(\ref{sumopdef.eq}). The
spin-isospin part can easily be evaluated in our spin and
isospin functions, Eq.~(\ref{spin.fn}), so that we focus on the 
matrix element of the spatial part. As is clear from Eqs.~(\ref{qtb.eq}) and 
(\ref{qtbdecom.eq}),
the spatial tensor operators have the form $[\bm a \times \bm b]_{\kappa
\mu}$, where $\bm a$ and $\bm b$ are vectors that represent one of 
the various coordinates, $\bm \rho_i, \ \bm \rho_j,\  \bm r_{ij},\  \bm
R_{ij}$. It is useful to note 
that any of these coordinates can be expressed as a 
linear combination of the relative coordinate set $\bm x$: 
$\bm a=\sum_{i=1}^{N-1}\omega_i\bm x_i=\tilde{\omega}{\bm x}$, and 
$\bm b=\sum_{i=1}^{N-1}\zeta_i\bm x_i=\tilde{\zeta}{\bm x}$, where  
$\omega$ and $\zeta$ are both $(N-1)$-dimensional column vectors.    
Therefore it is sufficient to show how we can evaluate 
the quadratic spatial tensor operators, 
$[\tilde{\omega}\bm{x}\times\tilde{\zeta}\bm{x}]_{\kappa\mu}$,   
with the basis functions~(\ref{GVR.eq}). A detailed method of evaluation 
is presented in Ref.~\cite{suzuki08}, and here we follow its 
formulation and notation.

First we calculate the matrix element between the generating
function 
\begin{align}
g(\bm{s},A,\bm{x})=\exp\Big(-\frac{1}{2}\tilde{\bm{x}}A\bm{x}+\tilde{\bm{s}}\bm{x}\Big),
\end{align}
where $\bm s$ is an $(N-1)$-dimensional column vector whose $i$th
element is a 3-dimensional vector $\bm s_i$, and $\tilde{\bm{s}}\bm{x}$
is a short-hand notation of $\sum_{i=1}^{N-1}\bm s_i\cdot{\bm x}_i$. 
As given in Ref.~\cite{svm}, it reads
\begin{widetext}
\begin{align}
& \left< g(\bm{s}^\prime,A^\prime, \bm{x}) \right|
[\tilde{\omega}\bm{x}\times\tilde{\zeta}\bm{x}]_{\kappa\mu}
\left| g(\bm{s},A,\bm{x}) \right> \notag\\
&=\left\{-\sqrt{3}\delta_{\kappa 0}\delta_{\mu 0}
{\rm Tr}\left(B^{-1}\omega\tilde{\zeta}\right)+
\left[\tilde{\omega}B^{-1}\bm{v}\times\tilde{\zeta}B^{-1}\bm{v}\right]_{\kappa\mu}
\right\}\left(\frac{(2\pi)^{N-1}}{{\rm det}B}\right)^{\frac{3}{2}}\text{e}^{\frac{1}{2}
\tilde{\bm{v}}B^{-1}\bm{v}}, 
\label{gene.eq}
\end{align}
\end{widetext}
where Tr stands for a trace and 
\begin{align}
B=A+A^{\prime}, \quad\bm{v}=\bm{s}+\bm{s}^\prime.
\end{align}
Using the $(N-1)$-dimensional
column vector $u_i$ specifying the GV
we express $\bm s$ and $\bm{s}^\prime$ as 
$\bm s=\lambda_1\bm e_1u_1+\lambda_2\bm e_2u_2$ and $\bm
s'=\lambda_3\bm e_3u_3+\lambda_4\bm e_4u_4$, where a unit 
vector $\bm e_i$ ($(\bm e_i\cdot\bm e_i)=1$) and a parameter $\lambda_i$ 
are introduced to manipulate the calculation of the sought matrix element.
See Ref.~\cite{suzuki08} for details.
The second term in the curly bracket and the exponential
function of Eq.~(\ref{gene.eq}) is simplified to
\begin{align}
&\left[\tilde{\omega}B^{-1}\bm{v}\times\tilde{\zeta}B^{-1}\bm{v}\right]_{\kappa\mu}
\to\sum_{i,j}f_ig_j
\lambda_i\lambda_j[\bm{e}_i\times\bm{e}_{j}]_{\kappa\mu},
\notag \\
&\text{e}^{\frac{1}{2}\tilde{\bm{v}}B^{-1}\bm{v}} \to \text{e}^{\sum_{i<j}\rho_{ij}\lambda_i\lambda_j\bm{e}_i\cdot\bm{e}_j},
\end{align}
where 
\begin{align}
\rho_{ij}=\tilde{u}_iB^{-1}u_j,\quad
f_i=\tilde{\omega}B^{-1}u_i, \quad g_j=\tilde{\zeta}B^{-1}u_j.
\end{align}
Here the arrow symbol indicates that both sides are equal as
long as the calculation of the sought matrix element is concerned. 
That is, any terms that have 
$\lambda_i^2(\bm e_i\cdot\bm e_i)=\lambda_i^2$ dependence make no
contribution to the matrix element, so that they can be dropped. \\

\par\noindent
(i) {\it $\kappa=0$ case}

In this case the term 
$[\tilde{\omega}B^{-1}\bm{v}\times\tilde{\zeta}B^{-1}\bm{v}]_{00}$
produces the same structure, with respect to 
$\lambda_i\lambda_j(\bm{e}_i\cdot\bm{e}_j)$, as the 
kinetic and mean square distance
operators. See Appendix B.2 of Ref.~\cite{suzuki08}.
The matrix element is
\begin{widetext}
\begin{align}
&\left<F_{(L_3L_4)LM}(u_3,u_4,A^\prime,\bm{x})\right|
[\tilde{\omega}\bm{x}\times\tilde{\zeta}\bm{x}]_{00}
\left|F_{(L_1L_2)LM}(u_1,u_2,A,\bm{x})\right>\notag \\
&=-\frac{1}{\sqrt{3}}\Bigl\{3{\rm Tr}(B^{-1}\omega\tilde{\zeta})
+\sum_{i<j}(f_ig_j+f_jg_i)
\frac{\partial}{\partial \rho_{ij}}\Bigr\}
\left<F_{(L_3L_4)LM}(u_3,u_4,A^\prime,\bm{x})|
F_{(L_1L_2)LM}(u_1,u_2,A,\bm{x})\right>.
\end{align}
\end{widetext}
Compare this expression with Eq.~(B.17)~\cite{suzuki08}. 
A formula for the overlap matrix 
element, $\langle F_{(L_3L_4)LM}(u_3,u_4,A^\prime,\bm{x})|
F_{(L_1L_2)LM}(u_1,u_2,A,\bm{x})\rangle$, is given in Eq.~(B.10)~\cite{suzuki08}. \\

\par\noindent
(ii) {\it $\kappa=1$ case}

The $\kappa=1$ case can be evaluated in exactly the
same way as the spin-orbit matrix element 
of Ref.~\cite{suzuki08}.  
The result is
\begin{widetext}
\begin{align}
&\left<F_{(L_3L_4)L^\prime M^\prime}(u_3,u_4,A^\prime,\bm{x})\right|
[\tilde{\omega}\bm{x}\times\tilde{\zeta}\bm{x}]_{1\mu}
\left|F_{(L_1L_2)LM}(u_1,u_2,A,\bm{x})\right>\notag\\
&=-\frac{4\pi}{\sqrt{3}}\frac{(-1)^{L_1+L_2+L+L^\prime}}{\sqrt{2L^\prime+1}}
\left<LM1\mu|L^\prime M^\prime\right>
\sum_{l>k=1}^{4}(f_kg_l-f_lg_k)(-1)^{\bar{L}_1+\bar{L}_2}
\left(\prod_{i=1}^{4}\frac{B_{L_i}}{B_{\bar{L}_i}}\right)\notag\\
&\times\sum_{\bar{L}}\sqrt{2\bar{L}+1}
Z_2(1\bar{L}_1\bar{L}_2\bar{L}_3\bar{L}_4\bar{L},LL^\prime;kl)
\left<F_{(\bar{L}_3\bar{L}_4)\bar{L}\bar{M}}(u_3,u_4,A^\prime,\bm{x})|
F_{(\bar{L}_1\bar{L}_2)\bar{L}\bar{M}}(u_1,u_2,A,\bm{x})\right>.
\end{align}
\end{widetext}
Compare this expression with Eq. (B.54)
~\cite{suzuki08}. 
The barred angular momentum labels  
$\bar{L}_i$ and $\bar{L^\prime}$ follow 
the definitions in Ref.~\cite{suzuki08}.
The coefficient $Z_2$ is defined in
Eq.~(B. 48)~\cite{suzuki08}.\\

\par\noindent
(iii) {\it $\kappa=2$ case}

In this case we note that
\begin{align}
&\left[\tilde{\omega}B^{-1}\bm{v}\times\tilde{\zeta}B^{-1}\bm{v}\right]_{2\mu}
\notag \\
&\to \sqrt{\frac{8\pi}{15}}\sum_{i=1}^4 f_ig_i\lambda_i^2Y_{2\mu}(\bm{e}_i)
\notag\\
&
+\frac{4\pi}{3}\sum_{i<j}
(f_ig_j+f_jg_i)\lambda_i\lambda_j
\left[\bm{e}_i\times\bm{e}_j\right]_{2\mu}.
\end{align}
Comparing this expression with Eqs. (B.41) and (B.42) 
and using Eq. (B.49)~\cite{suzuki08}, 
we obtain the matrix element as follows:
\begin{widetext}
\begin{align}
&\left<F_{(L_3L_4)L^\prime M^\prime}(u_3,u_4,A^\prime,\bm{x})\right|
[\tilde{\omega}\bm{x}\times\tilde{\zeta}\bm{x}]_{2\mu}
\left|F_{(L_1L_2)LM}(u_1,u_2,A,\bm{x})\right>\notag\\
&=\frac{(-1)^{L_1+L_2+L+L^\prime}}{\sqrt{2L^\prime+1}}
\left<LM2\mu|L^\prime M^\prime\right>\sqrt{5}
\left\{\sqrt{\frac{8\pi}{15}}
\sum_{k=1}^4 f_kg_k(-1)^{\bar{L}_1+\bar{L}_2}
\left(\prod_{i=1}^{4}\frac{B_{L_i}}{B_{\bar{L}_i}}\right)\right.\notag\\
&\times\sum_{\bar{L}}\sqrt{2\bar{L}+1}
Z_1(2\bar{L}_1\bar{L}_2\bar{L}_3\bar{L}_4\bar{L},LL^\prime;k)
\left<F_{(\bar{L}_3\bar{L}_4)\bar{L}\bar{M}}(u_3,u_4,A^\prime,\bm{x})|
F_{(\bar{L}_1\bar{L}_2)\bar{L}\bar{M}}(u_1,u_2,A,\bm{x})\right>\notag\\
&+\frac{4\pi}{3}\sum_{l>k=1}^{4}(f_kg_l+f_lg_k)(-1)^{\bar{L}_1+\bar{L}_2}
\left(\prod_{i=1}^{4}\frac{B_{L_i}}{B_{\bar{L}_i}}\right)
\notag\\
&\times
\sum_{\bar{L}}\sqrt{2\bar{L}+1}
Z_2(2\bar{L}_1\bar{L}_2\bar{L}_3\bar{L}_4\bar{L},LL^\prime;kl)
\Bigg.\left<F_{(\bar{L}_3\bar{L}_4)\bar{L}\bar{M}}(u_3,u_4,A^\prime,\bm{x})|
F_{(\bar{L}_1\bar{L}_2)\bar{L}\bar{M}}(u_1,u_2,A,\bm{x})\right>\Bigg\}.
\end{align}
\end{widetext}
The coefficient $Z_1$ is defined in Eq.~(B. 46)~\cite{suzuki08}.

\section{Contribution of the kinetic energy to the spin-dipole energy-weighted
 sum rule}

The aim of this appendix is to derive Eq.~(\ref{sd.ewsr.formula}). 
Introducing an abbreviation 
\begin{align}
v_{\lambda \mu}(i)=[\bm \rho_i \times \bm \sigma_i]_{\lambda \mu}
\end{align}
and $T=\sum_{i=1}^AT_i-T_{\rm cm}$, 
we calculate $X^p_{(\lambda)0}(T)$ from the following expression 
\begin{align}
X^p_{(\lambda)0}(T)&=\frac{1}{2}\sum_{\mu}\sum_{i,j=1}^A
\Big[ v_{\lambda \mu}^{\dagger}(j){T^p_{j}}^{\dagger}, \big[T, v_{\lambda \mu}(i)\big]T^p_i  \Big]  \notag \\
&=\frac{1}{2}\sum_{\mu}\sum_{i,j=1}^A
\Big\{v_{\lambda \mu}^{\dagger}(j)[T,v_{\lambda
 \mu}(i)][{T^p_{j}}^{\dagger}, T^p_i] \notag \\
&+\Big[v_{\lambda \mu}^{\dagger}(j),[T,v_{\lambda
 \mu}(i)]\Big]T^p_i{T^p_{j}}^{\dagger}\Big\}.
\label{formula.X.kine}
\end{align}
Here use is made of the relation $[AB,CD]=AC[B,D]+[A,C]DB$ provided that
$[A,D]=0$ and $[B,C]=0$.
The first term in the curly bracket is
contributed only by $i=j$ terms because 
$[{T^p_{j}}^{\dagger}, T^p_i]$ vanishes for $i\neq j$.  Using
the commutation relation 
\begin{align}
[T,v_{\lambda \mu}(i)]
=-\frac{i\hbar}{m_N} [(\bm
 p_i-\textstyle{\frac{1}{A}}\bm P_{\rm tot})\times \bm \sigma_i]_{\lambda \mu}, 
\end{align}  
we obtain the first term as 
\begin{align}
&{\rm First \ term}=-\frac{i\hbar }{2m_N}(-1)^{\lambda}\notag \\
&\times \sum_{i=1}^A 
\Big([\bm \rho_i \times \bm \sigma_i]_{\lambda}\cdot [(\bm p_i-\textstyle{\frac{1}{A}}\bm P_{\rm
 tot})\times \bm \sigma_i]_{\lambda}\Big)
[{T^p_{i}}^{\dagger}, T^p_i]. 
\end{align}
The ground-state expectation value of this term is conveniently 
evaluated by decomposing the above scalar product to that of 
the space-space and spin-spin terms using the matrix $U$ of 
Eq.~(\ref{ucoef.eq}). The result is 
\begin{align}
&{\rm First \ term}=-\frac{i\hbar U_{\lambda\, 0}}{2m_N}\sum_{i=1}^A
\big(\bm \rho_i\cdot(\bm p_i-\textstyle{\frac{1}{A}}\bm P_{\rm tot})\big)[{T^p_{i}}^{\dagger}, T^p_i]\notag \\
&-\frac{\hbar U_{\lambda\, 1}}{2m_N}\sum_{i=1}^A
\big((\bm \rho_i\times(\bm p_i-\textstyle{\frac{1}{A}}\bm P_{\rm
 tot}))\cdot\bm \sigma_i\big)[{T^p_{i}}^{\dagger}, T^p_i].
\label{firstterm}
\end{align}
The matrix element of the spatial part involving the operators,  
$\big(\bm \rho_i\cdot(\bm p_i-\textstyle{\frac{1}{A}}\bm P_{\rm
tot})\big)$ 
and $\big((\bm \rho_i\times(\bm p_i-\textstyle{\frac{1}{A}}\bm P_{\rm
 tot}))\cdot\bm \sigma_i\big)$, can be calculated in the manner similar
 to that presented in Appendix B. See Ref.~\cite{suzuki08} for the details. 

The second term in the curly bracket of Eq.~(\ref{formula.X.kine}) can
be obtained in a similar way. 
After a straightforward calculation of the commutator, we
obtain the following result:
\begin{align}
&{\rm Second \
 term}\notag \\
&=\frac{\hbar^2}{2m_N}(2\lambda+1)N^p-\frac{\hbar^2}{6Am_N}(2\lambda+1)
({\bm \Sigma}^p\cdot {{\bm \Sigma}^p}^{\dagger}) \notag \\ 
&+\frac{\hbar}{6m_N}C_{\lambda}\sum_{i=1}^A
\big((\bm \rho_i \times (\bm p_i-\textstyle{\frac{1}{A}}\bm P_{\rm tot}))
 \cdot \bm \sigma_i\big)T_i^p{T_i^p}^{\dagger},
\label{secondterm}
\end{align}
where $C_{\lambda}$ is 
\begin{align}
C_0=2,\quad C_1=3,\quad C_2=-5. 
\end{align}
Here the operators $N^p$ and $\bm \Sigma^p$ are defined by 
\begin{align}
N^p= \sum_{i=1}^AT_i^p{T^p_i}^{\dagger}, \quad 
\bm \Sigma^p=\sum_{i=1}^A \bm \sigma_iT_i^p,
\end{align}
which leads to $({\bm \Sigma}^p\cdot {{\bm
 \Sigma}^p}^{\dagger})=3N^p+\sum_{j>i=1}^A(\bm \sigma_i\cdot\bm
 \sigma_j)T^p_{ij}$.
Combining (\ref{firstterm}) and
 (\ref{secondterm}) we obtain Eq.~(\ref{sd.ewsr.formula}).


\begin{thebibliography}{99}
\bibitem{gazit07} D. Gazit and N. Barnea, Phys. Rev. Lett. {\bf 98}, 192501 (2007).
\bibitem{tsuzuki06} T. Suzuki, S. Chiba, T. Yoshida, T. Kajino, and T. Otsuka,
Phys. Rev. C {\bf 74}, 034307 (2006).
\bibitem{fujita11} Y. Fujita, B. Rubio, and W. Gelletly,
Prog. Part. Nucl. Phys. {\bf 66}, 549 (2011).
\bibitem{okamura02}H. Okamura {\it et al.}, Phys. Rev. C {\bf
		66}, 054602 (2002).
\bibitem{huu07} M.~A. de Huu {\it et al.}, Phys. Lett. {\bf B\,649}, 35 (2007).
\bibitem{gaarde84} C. Gaarde {\it et al.}, Nucl. Phys. {\bf A\,422}, 189 (1984). 
\bibitem{rapaport94} J. Rapaport and E. Sugerbaker, Annu. Rev. Nucl. Part. Sci. {\bf 44}, 109 (1994).
\bibitem{nakayama07} S. Nakayama {\it et al.}, Phys. Rev. C {\bf 76}, 021305(R) (2007).
\bibitem{nakayama08} S. Nakayama {\it et al.}, Phys. Rev. C {\bf 78}, 014303 (2008).
\bibitem{wakasa11} T. Wakasa {\it et al.}, Phys. Rev. C {\bf 84}, 014614 (2011).
\bibitem{dumitrescu84} T.~S. Dumitrescu and T. Suzuki,
Nucl. Phys. {\bf A\,423}, 277 (1984).
\bibitem{tsuzuki98} T. Suzuki and H. Sagawa, Nucl. Phys. {\bf A\,637}, 547 (1998).
\bibitem{bai10} C.~L. Bai, H.~Q. Zhang, H. Sagawa, X.~Z. Zhang, G. Col\'{o}, 
and F.~R. Xu, Phys. Rev. Lett. {\bf 105}, 072501 (2010).
\bibitem{bai11} C.~L. Bai, H. Sagawa, G. Col\'{o}, H.~Q. Zhang, and X.~Z. Zhang,
Phys. Rev. C {\bf 84}, 044329 (2011).
\bibitem{liang12} H. Liang, P. Zhao, and J. Meng, Phys. Rev. C {\bf 85}, 064302 (2012).
\bibitem{kamada01} H. Kamada {\it et al.}, Phys. Rev. C {\bf 64}, 044001 (2001).
\bibitem{forest96} J.~L. Forest, V.~R. Pandharipande, S.~C. Pieper, R.~B. Wiringa, R. Schiavilla, and A. Arriaga, Phys. Rev. C {\bf 54}, 646 (1996).
\bibitem{feldmeier11} H. Feldmeier, W. Horiuchi, T. Neff, and Y. Suzuki,
Phys. Rev. C {\bf 84}, 054003 (2011).
\bibitem{horiuchi12a} W. Horiuchi, Y. Suzuki, and K. Arai, Phys. Rev. C {\bf 85},
054002 (2012).
\bibitem{horiuchi08} W. Horiuchi and Y. Suzuki, Phys. Rev. C {\bf 78},
	034305 (2008).
\bibitem{horiuchi12b} W. Horiuchi and Y. Suzuki, Few-Body Syst., in
	press, DOI 10.1007/s00601-012-0495-y.
\bibitem{ho83} Y.~K. Ho, Phys. Rep. {\bf 99}, 1 (1983).
\bibitem{moiseyev98} N. Moiseyev, Phys. Rep. {\bf 302}, 211 (1998).
\bibitem{CSM} S. Aoyama, T. Myo, K. Kat\={o}, and K. Ikeda, 
	Prog. Theor. Phys. {\bf 116}, 1 (2006).
\bibitem{AV8p} B.~S. Pudliner, V.~R. Pandharipande, J. Carlson,
	S.~C. Pieper, and R.~B. Wiringa, Phys. Rev. C {\bf 56}, 1720 (1997).
\bibitem{G3RS} R. Tamagaki, Prog. Theor. Phys. {\bf 39}, 91 
	(1968).
\bibitem{hiyama04} E. Hiyama, B.~F. Gibson, and M. Kamimura, 
	Phys. Rev. C {\bf 70}, 031001(R) (2004).
\bibitem{boys60} S.~F. Boys, Proc. R. Soc. London Ser. A {\bf 258}, 402 (1960).
\bibitem{singer60} K. Singer, Proc. R. Soc. London Ser. A {\bf 258}, 412 (1960).
\bibitem{varga97}K. Varga, Y. Ohbayasi, and Y. Suzuki, Phys. Lett. {\bf	B\,396}, 1 (1997).
\bibitem{mitroy13} J. Mitroy {\it et al.}, Rev. Mod. Phys., in press.
\bibitem{varga95}K. Varga and Y. Suzuki, Phys. Rev. C {\bf 52}, 2885 (1995).
\bibitem{svm} Y. Suzuki and K. Varga, {\it Stochastic Variational Approach to
	Quantum-Mechanical Few-Body Problems}, Lecture Notes in Physics,
	(Springer, Berlin, 1998), Vol. m54.
\bibitem{suzuki08} Y. Suzuki, W. Horiuchi, M. Orabi, and K. Arai, 
Few-Body Syst. {\bf 42}, 33 (2008).
\bibitem{aoyama12} S. Aoyama, K. Arai, Y. Suzuki, P. Descouvemont, and D. Baye,
Few-Body Syst. {\bf 52}, 97 (2012).
\bibitem{suzuki00}  
Y. Suzuki and J. Usukura, Nucl. Inst. Meth. {\bf B\,171}, 67 (2000).
\bibitem{varga94}K. Varga, Y. Suzuki, and R.~G. Lovas, Nucl. Phys. {\bf
	A\,571}, 447 (1994).
\bibitem{tilley92} D.~R. Tilley, H.~R. Weller, and G.~M. Hale,
Nucl. Phys. {\bf A\,541}, 1 (1992).
\bibitem{woosely90} S.~E. Woosley, D.~H. Hartmann, R.~D. Hoffman, and
	W.~C. Haxton, Astrophys. J. {\bf 356}, 272 (1990).
\bibitem{horiuchiNIC} W. Horiuchi, Y. Suzuki, and T. Sato,
Proc. of Science, PoS (NIC XI) 150 (2011).
\bibitem{smith60}
F.~T. Smith, Phys. Rev. {\bf 118}, 349 (1960).
\bibitem{igarashi04} A. Igarashi and I. Shimamura, Phys. Rev. A {\bf 70}, 
012706 (2004).
\bibitem{usukura02} J. Usukura and Y. Suzuki, Phys. Rev. A {\bf 66}, 010502 (R) (2002).  
\bibitem{suzuki04}Y. Suzuki and J. Usukura, Nucl. Inst. Meth. {\bf B\,221}, 195 (2004).
\bibitem{hazi70} A.~U. Hazi and H.~S. Taylor, Phys. Rev. A {\bf 1}, 1109 (1970). 
\bibitem{lipparini89} E. Lipparini and S. Stringari, 
Phys. Rep. {\bf 175}, 103 (1989).
\bibitem{tsuzuki84} T. Suzuki, Ann. Phys. Fr. {\bf 9}, 535 (1984).
\bibitem{MN} D.~R. Thompson, M. LeMere, and Y.~C. Tang, Nucl. Phys. {\bf A\,286}, 53 (1977). 
\bibitem{gaarde81} C. Gaarde {\it et al.}, 
Nucl. Phys. {\bf A\,369}, 258 (1981).
\bibitem{tsuzuki79} T. Suzuki, Phys. Lett. {\bf 83B}, 147 (1979).
\end{thebibliography}
\end{document}